\documentstyle[preprint,pre,eqsecnum,aps]{revtex}
\begin{document}
\draft

\date{\today}
\title{Theoretical Study of Fluid Membranes of Spherical Topology with
Internal Degrees of Freedom.}
\author{R. M. L. Evans}
\address{Theoretical Physics Group \\
Department of Physics and Astronomy \\
The University of Manchester, M13 9PL, UK}
\maketitle

\begin{abstract}
  A theoretical study of vesicles of topological genus zero is presented. The
bilayer membranes forming the vesicles have various degrees of intrinsic
(tangent-plane) orientational order, ranging from smectic to hexatic,
frustrated by curvature and topology. The field-theoretical model for
these `$n$-atic' surfaces has been studied before in the low temperature
(mean-field) limit. Work presented here includes the effects of thermal
fluctuations. Using the lowest Landau level approximation, the coupling
between order and shape is cast in a simple form, facilitating insights into
the behaviour of vesicles. The order parameter contains vortices, whose
effective interaction potential is found, and renormalized by membrane
fluctuations. The shape of the phase space has a counter-intuitive influence on
this potential. A criterion is established whereby a vesicle
of finite rigidity may be burst by its own in-plane order, and an analogy is
drawn with flux exclusion from a type-I superconductor.
\end{abstract}


\section{Introduction}

  When amphiphilic molecules (those with hydrophobic and hydrophilic parts) are
dissolved in water, they hide their aliphatic tails by grouping together and
thus assemble themselves into structures. Depending on the geometry and
chemistry of the particular surfactant, these structures may be anything from
nanometre-scale mycils to macroscopic membranes arranged in stacks,
bicontinuous networks of pipes, or closed surfaces. The literature on
spontaneously self-assembling amphiphiles and biological and liquid-crystal
membranes is vast. See for example \cite{Lipowsky91}, \cite{David89} or
\cite{Peliti94}.

  Research presented here is concerned with bilayer fluid
membranes with internal orientational degrees of freedom. Fluidity of the
membrane refers to the fact that molecules can move freely {\em within} the
surface, although {\em perpendicular} deformations have an energy cost. The
molecules have no positional crystalline order as the temperature is above the
crystalline-to-fluid transition at which lattice dislocations proliferate.
However, for some liquid crystals, this is not coincident with loss of {\em
orientational} order, which occurs at a higher temperature. Various kinds of
orientational order are possible. Liquid crystals in the smectic-C phase have
carbon-chain tails which are tilted with respect to the local normal to the
surface. The local mean direction of tilt, projected onto the local tangent
plane of the membrane, gives a two-component vector order parameter, which
disappears at the continuous transition to smectic-A phase, where rotational
symmetry is restored \cite{MacKintosh91}. Alternatively, the orientational
order may be hexatic. In this case, the Kosterlitz-Thouless (K-T) transition at
which lattice disclinations proliferate, destroying the quasi-long-range
hexagonal bond orientational order, is non-coincident with the lower
temperature K-T transition at which dislocations proliferate \cite{Nelson79}.
Nearest neighbour bond orientations give a basis for defining an order
parameter, which again has two components in the local tangent plane. But a
six-fold ambiguity in the choice of nearest neighbours arises from the local
six-fold rotational invariance of the membrane. Mapping the local tangent plane
onto the Argand plane, with some arbitrary direction chosen for the real axis,
allows the two-component order parameter to be well defined as a complex
number thus:
\begin{displaymath}
  \psi(\mbox{\boldmath $\sigma$}) =
  \left< \exp ( i 6 \Theta(\mbox{\boldmath $\sigma$}) ) \right> ,
\end{displaymath}
where {\boldmath $\sigma$} is some two-component coordinate defined on the
surface, and $\Theta$ is the bond angle with respect to an arbitrary coordinate
axis. Notice that this definition is independent of which of the six bonds is
chosen, since rotations of the membrane through $\pi/3$ correspond to a phase
change of $2\pi$ in $\psi$, or an identity transformation. Hence, this
two-component order parameter has the appropriate six-fold local rotational
invariance. A similar complex order parameter can be defined for the smectic-C
order described above. This has local full-turn invariance only, and magnitude
$\psi_{0}$. Hence it may be mapped onto a complex field thus:
\begin{displaymath}
  \psi_{\mbox{c}}(\mbox{\boldmath $\sigma$}) =
  \left< \psi_{0} \exp ( i \Theta(\mbox{\boldmath $\sigma$}) ) \right> ,
\end{displaymath}
where $\Theta$ gives the orientation of the tail vectors projected onto the
tangent plane. In each case, $\left<\right>$ denotes a local thermal average.
The notion
may be generalized to an in-plane orientational order parameter with local
$n$-fold rotational invariance being represented by a complex field
\begin{equation}
\label{psi}
  \psi(\mbox{\boldmath $\sigma$}) =
  \left< \psi_{0} \exp ( i n \Theta(\mbox{\boldmath $\sigma$}) ) \right> ,
\end{equation}
where $\Theta$ gives the orientation of one of the $n$ principal directions of
order. For smectic and bond-angle order in liquid crystals, $n\in\{1,2,4,6\}$.
Geometrical arguments for this restriction on the possible orders of rotational
symmetry are given in \cite{Evans95}.

  On a flat membrane, the continuous development of $\psi$ is well modelled by
a Ginzburg-Landau free energy functional of the form
\begin{equation}
\label{prelim}
  {\cal H} = \int \left\{ r |\psi|^2 + \mbox{$\frac{1}{2}$} u |\psi|^4 +
  C |\nabla\psi|^2 \right\}\, d\cal{A} ,
\end{equation}
where $d\cal{A}$ is an element of area, and $r$, $u$ and $C$ are parameters of
the particular conditions of the system being modelled.

  On a flexible membrane however, the free energy functional is a little more
complicated. Let the membrane be treated as an ideal, two-dimensional, smooth
surface. A Hamiltonian due to Helfrich \cite{Helfrich86} governs elastic
deformations of an isotropic ($\psi=0$) fluid membrane. This Hamiltonian is a
functional of the intrinsic and extrinsic curvatures of the membrane, as
defined below. Let $\mbox{\boldmath $R$} (\mbox{\boldmath $\sigma$})$ be the
position vector in $R^3$ of a point on the surface with two-dimensional
coordinate $\mbox{\boldmath $\sigma$}=(\sigma_{1},\sigma_{2})$. The surface has
a metric tensor with components $g_{ab}=\partial_{a}\mbox{\boldmath
$R$}\cdot\partial_{b}\mbox{\boldmath $R$}$ where $a$ and $b$ label the
coordinates. The metric has inverse $g^{ab}$ and determinant $g$. A covariant
curvature tensor is defined by $K_{ab}=\mbox{\boldmath $N$}\cdot \partial_{a}
\partial_{b} \mbox{\boldmath $R$}$ where $\mbox{\boldmath $N$}$ is the local
unit normal. Note that dot products are evaluated in $R^3$. Using Einstein
summation convention, and the metric and its inverse to lower and raise indices
in the standard way, a more useful curvature tensor is defined thus:
$K^a_{b}=g^{ac}K_{cb}$. Its determinant $K$ is the Gaussian curvature of the
surface, also called the intrinsic curvature, as it is experienced by
`flatlanders' and other physical entities living within the two-dimensional
world of the surface, who are oblivious to the way in which the surface is
embedded in 3-space. Its trace $K^a_{a}$ however is an extrinsic property,
which cannot be determined by flatlanders\footnote{The idea of `flatlanders'
was used by E. A. Abbott in his book ``Flatland"\cite{flat}. They are
hypothetical beings who live in a two-dimensional space, and have no
conception of a third dimension.}. This quantity is known as the total
curvature, or twice the mean curvature, and is the sum of the two principal
radii of curvature of the surface at a given point \cite{Rutherford59}.
Helfrich's Hamiltonian, describing the elastic properties of a constant area
fluid membrane, can now be expressed as
\begin{equation}
\label{Helfrich}
  {\cal H}_{\mbox{{\small Helfrich}}}
  = \int\{\mbox{$\frac{1}{2}$}\kappa(K^a_{a})^2 +
  \kappa_{G}K\} \sqrt{g}\,d^2\sigma.
\end{equation}
Note that $\sqrt{g}\,d^2\sigma=d{\cal A}$. The phenomenological constants in
this expression are the bending rigidity $\kappa$ and the Gaussian curvature
modulus $\kappa_{G}$. On a closed surface, the second term in equation
(\ref{Helfrich}) is a topological invariant, according to the Gauss-Bonnet
formula
\begin{displaymath}
  \int K\,d{\cal A} = 2\pi\chi
\end{displaymath}
where $\chi$ is the Euler number of the surface \cite{Weeks85}. For an
orientable surface of genus (number of `handles') G, the Euler number is given
by
\begin{displaymath}
  \chi=2(1-G).
\end{displaymath}
Hence the dynamics of a membrane of given topology are not influenced by the
second term of equation (\ref{Helfrich}). What follows will concentrate on
membranes of spherical topology (genus zero), so this term will be dropped.

  The free energy functional governing the dynamics of $n$-atic fluid membranes
cannot be found by simply adding together equations (\ref{prelim}) and
(\ref{Helfrich}), since (\ref{prelim}) contains a gradient operator, which must
be expressed in terms of coordinates which are no longer flat, but curvilinear.
The derivatives in the flat-membrane expression \mbox{$|\nabla\psi|^2 =
\partial_{x}\psi^*\partial_{x}\psi+\partial_{y}\psi^*\partial_{y}\psi$} must be
replaced by covariant derivatives thus: \mbox{$|\nabla\psi|^2 = D_{a}^*\psi^*
D^a\psi$} where $D_{a}$ is of the form $\partial_{a}-inA_{a}$ with the
connection $\mbox{\boldmath $A$}$ defined such that the free energy is
invariant under general coordinate transformations. Clearly, $\mbox{\boldmath
$A$}$ itself is not coordinate-invariant, since $\psi$ has coordinate
dependence due to the arbitrariness of the reference axis for $\Theta$ in
equation (\ref{psi}). There is some freedom in the choice of the gauge field
$\mbox{\boldmath $A$}$. Here, $\mbox{\boldmath $A$}$ is chosen to be the `spin
connection' \mbox{$A_{a}=\mbox{\boldmath $e$}_{1} \cdot
\partial_{a}\mbox{\boldmath $e$}_{2}$} where $\mbox{\boldmath $e$}_{b}$ is a
unit vector in the direction of $\partial_{b}\mbox{\boldmath $R$}$. Finally, we
have the free energy functional for a closed, $n$-atic fluid membrane:
\begin{equation}
\label{F}
  F[\psi(\mbox{\boldmath $\sigma$}),\mbox{\boldmath $R$}
  (\mbox{\boldmath $\sigma$})]=\int\,d^2\sigma\sqrt{g} \left\{ r |\psi|^2 +
  \mbox{$\frac{1}{2}$} u |\psi|^4 + C D_{a}^*\psi^*D^{a}\psi +
  \mbox{$\frac{1}{2}$}\kappa(K^a_{a})^2 \right\}
\end{equation}
and, since the membrane's total area is fixed at ${\cal A}$ by the number of
incompressibly packed constituent molecules, there is also a constraint
\begin{equation}
\label{area}
  \int\,d^2\sigma\sqrt{g}={\cal A}.
\end{equation}
The covariant derivatives provide a coupling between order and the shape of the
membrane.

  So far, the model and formalism are identical to that used by Park, Lubensky
and MacKintosh in their mean-field treatment of $n$-atic fluid membranes of
genus-zero (vesicles) \cite{Park92}. They noted that, not only is $n$-atic
ordering partly frustrated by curvature of its two-dimensional space, making
parallelism impossible to achieve, but it is further frustrated by the
spherical boundary conditions. Orientational order intrinsic to a surface of
Euler number $\chi$ must have topological defects whose indices sum to $\chi$.
For instance, smectic-C (vector) order can form defects (sometimes called
vortices) of index 1. Hence, on a sphere (Euler number 2) and in the absence of
any anti-vortices (index -1 defects), there must be two defects in a Sm-C
order parameter. At these poles, $\psi$ vanishes smoothly to avoid infinite
gradients. Clearly, an $n$-atic can form defects of index $1/n$ since it can
rotate by less than a full turn in circling a pole, without its phase slipping.
So $n$-atics form $2n$ vortices on a sphere, in the absence of excited
vortex-anti-vortex pairs.

  A similar approach to that of Park et al.\ will be used in this paper. The
small deformations from sphericity of the $n$-atic fluid membrane are expressed
as a real, scalar field, and the order parameter as a complex field $\psi$ as
described above. The shape deformation field is expanded in spherical
harmonics, and $\psi$ is expanded in eigenfunctions of the gradient operator.
As an approximation, the series expansion for $\psi$ is truncated and only the
degenerate set of lowest-eigenvalue functions are kept. The same approximation
was used by Landau in finding the wavefunction of an electron in a magnetic
field \cite{Landau58}, and by Abrikosov in his treatment of flux lattices in
superconductors \cite{Abrikosov}.
The study diverges from that of Park et al.\ when a simple expression is
calculated for the coupling between the two fields. The free energy is cast
into a form which allows the shape fluctuations to be integrated out exactly,
giving renormalized coefficients for the order parameter. This effectively maps
the deformable sphere problem onto a rigid sphere problem, which is soluble for
the $n=1$ case. In section \ref{stability}, the stability of the system is
analyzed to ascertain the conditions under which topological defects become too
expensive, and are excluded from the membrane in a similar fashion to flux
exclusion from a type-I superconductor.
The partition function and various expectation values for $n=1$
are calculated in section \ref{n=1}, using some special symmetries of the
problem. For other values of $n$, the system is not soluble, but approximate
expectation values are found in section \ref{H-F}, using diagrammatic expansion
and the Hartree-Fock method, which gives correct results in the limit of high
temperature.

\section{Calculation}

  Using the expression $D_{a}=\partial_{a}-inA_{a}$, it is easy to show that,
on a closed surface,
\begin{displaymath}
  \int\! d^2\sigma\sqrt{g}D^a\psi D^*_{a}\psi^*
  \equiv-\int\! d^2\sigma\,\psi^*D_{a}(\sqrt{g}D^a\psi).
\end{displaymath}
Both sides are real for all $\psi$ hence the operator
$-g^{-\frac{1}{2}} D_{a}(\sqrt{g}D^a)$
is Hermitian. Eigenvalues of this operator scale as ${\cal A}$. A useful set of
orthonormal basis functions is formed by normalized eigenfunctions of this
operator on the unit sphere,
where $\sqrt{g}\,d^2\sigma=d\Omega$ which is the usual solid angle element
($\sin\theta\,d\theta\,d\varphi$ in spherical polar coordinates).
These functions
are sometimes referred to as `Landau levels'. There are $(2n+1)$ degenerate
Landau levels with the lowest eigenvalue, $n$. If $\psi$ is expanded in this
complete set of functions then, close to the mean-field phase transition, the
partition function is dominated by configurations involving only the lowest
Landau levels. Hence, in this regime it is a good approximation to expand
$\psi$ in these complex, degenerate lowest levels only. Thus the number of
degrees of freedom in $\psi$ is reduced to $2(2n+1)$, corresponding to freedom
in the positions of the $2n$ vortices in the two-dimensional surface, together
with an overall complex amplitude. On the unit sphere,
\begin{equation}
\label{operator}
  -\frac{D_{a}(\sqrt{g}D^a)}{\sqrt{g}}=-(\partial_{\theta}\partial_{\theta}
  +\mbox{\,cosec}^2\theta\,\partial_{\varphi}\partial_{\varphi}
  +\cot\theta\,\partial_{\theta}
  +2in\mbox{\,cosec\,}\theta\cot\theta\,\partial_{\varphi}-n^2\cot^2\theta).
\end{equation}
The lowest eigenfunctions of this operator and similar operators for other
gauges, are presented in different forms in \cite{Roy83}, \cite{ONeill93} and
\cite{Park92}, and are re-expressed here as
\begin{equation}
\label{basis}
  \phi_{p}=\sqrt{\frac{(2n+1)!}{4\pi(n+p)!(n-p)!}}\sin^{n+p}(\frac{\theta}{2})
  \cos^{n-p}(\frac{\theta}{2})\exp(ip\varphi)
\end{equation}
for integer values of $p$ between $-n$ and $n$. Any function $\psi$ which is a
linear combination of these functions, obeys the relation
\begin{equation}
\label{relation}
  \partial_{\theta}\psi =
  (n\cot\theta-i\mbox{\,cosec\,}\theta\,\partial_{\varphi})\psi
\end{equation}

  Let deviations from sphericity of the vesicle's shape be parameterized by a
real, dimensionless, scalar field $\rho(\mbox{\boldmath $\sigma$})$ in the
following way. If the ground-state vesicle shape is a sphere of radius $R_{0}$
then, with the origin at the centre of the vesicle,  
\begin{equation}
\label{rho}
  |\mbox{\boldmath $R$}(\mbox{\boldmath $\sigma$})|
  = \left(1+\rho(\mbox{\boldmath $\sigma$})\right)\,R_{0} .
\end{equation}
This is a `normal gauge' parameterization, and therefore carries the correct
weight in a statistical ensemble \cite{Cai94}.

  It will emerge that $<\rho^2>\sim<|\psi|^4>$. Equation (\ref{F}) approximates
the $\psi$ potential by a series expansion, truncated at fourth order. Hence,
it is consistent to also truncate to order $\rho^2$ and $|\psi|^2\rho$. {\em
ie.} the vesicles under consideration deviate little from spheres. To this
order, $\sqrt{g}|\psi|^2=R_{0}^2\sin\theta(1+2\rho)|\psi|^2$ and
\begin{displaymath}
  -\psi^*D_{a}(\sqrt{g}D^a)\psi=n|\psi|^2\sin\theta
  +2n\psi^*\{i\partial_{\varphi}\rho\,\partial_{\theta}\psi
  -i\partial_{\theta}\rho
  \,\partial_{\varphi}\psi+2n\cos\theta\,\psi\,\partial_{\theta}\rho\} .
\end{displaymath}
Applying the relation (\ref{relation}) and integrating by parts gives
\begin{displaymath}
  \int-\psi^*D_{a}(\sqrt{g}D^a)\psi\,d^2\sigma
  =n\int d\Omega |\psi|^2(1-\nabla^2_{\perp}\rho) +{\cal O}(|\psi|^2\rho^2).
\end{displaymath}
where \mbox{$\nabla^2_{\perp}=(\partial_{\theta}\partial_{\theta} +
\mbox{\,cosec}^2\theta\,\partial_{\varphi}\partial_{\varphi} +
\cot\theta\,\partial_{\theta})$} is the covariant Laplacian. So the expression
for the free energy becomes
\begin{displaymath}
  F=R_{0}^2 \int d\Omega \left\{(r-r_{c})|\psi|^2(1+2\rho)
  +\mbox{$\frac{1}{2}$} u|\psi|^4 + r_{c}|\psi|^2 (2+\nabla^2_{\perp})\rho
  \right\}
  + {\cal H}_{\mbox{\small Helfrich}} + {\cal O}(|\psi|^6)
\end{displaymath}
where
\begin{displaymath}
  r_{c}\equiv -\frac{Cn}{R_{0}^2}.
\end{displaymath}
It also emerges that $|\psi|^2\sim(r-r_{c})$ below the mean-field transition.
See section \ref{valid} for more explanation.
Furthermore, from the definitions of $K^a_{a}$ and $\rho$, it follows that
\begin{displaymath}
  K^a_{a}=-\frac{2}{R_{0}}(1-\rho-\mbox{$\frac{1}{2}$}\nabla^2_{\perp}\rho
  +\rho^2+\rho\nabla^2_{\perp}\rho)
\end{displaymath}
and, from equation (\ref{area}),
\begin{displaymath}
  {\cal A}=4\pi R_{0}^2 + R_{0}^2 \int d\Omega\,\rho(1+\mbox{$\frac{1}{2}$}
  \nabla^2_{\perp})\rho
\end{displaymath}
since $\rho$ and $R_{0}$ are defined in equation (\ref{rho}) in such a way that
$\int\rho\,d\Omega\equiv 0$.

  Hence, to ${\cal O}(|\psi|^4)$, we may write
\begin{displaymath}
  r_{c}=-\frac{4\pi nC}{\cal A}
\end{displaymath}
and the free energy
\begin{equation}
\label{simple}
  F[\psi,\rho]=\int d\Omega\left[\frac{\cal A}{4\pi}\left\{ (r-r_{c})|\psi|^2
  +\mbox{$\frac{1}{2}$} u|\psi|^4\right\}
  - Cn|\psi|^2(2+\nabla^2_{\perp})\rho +\kappa\left\{
  2+\rho\nabla^2_{\perp}\rho+\mbox{$\frac{1}{2}$}(\nabla^2_{\perp}\rho)^2
  \right\}\right]
\end{equation}
with the field $\rho$ unaffected by the fixed area constraint to this order.

 The coupling between the fields $\psi$ and $\rho$, describing order and shape,
has been cast into a very simple form, using some special properties of the
lowest Landau levels of $\psi$ in the spherical geometry. Note that a similar
restriction has not been put on the phase space of $\rho$. All modes of shape
fluctuation are available for small amplitude excitation. The remarkable
simplicity of equation (\ref{simple}) is conducive to further exploration of
the properties of $n$-atic vesicles, hitherto hindered by unwieldy notations.

  It is clear from equation (\ref{simple}) that the mean-field transition
temperature has been shifted from $r=0$ to the lower temperature $r=r_{c}$ due
to the non-zero lowest eigenvalue of the operator in equation (\ref{operator})
in the spherical geometry. Hence, ordering is frustrated by the curvature of
the space.

  Now let $\rho$ be expanded in eigenfunctions of the covariant Laplacian
$\nabla^2_{\perp}$. These functions are spherical harmonics $Y_{l}^m$, defined
by
\begin{displaymath}
  \nabla^2_{\perp} Y_{l}^m = -l(l+1) Y_{l}^m .
\end{displaymath}
The s-wave ($l=0$) harmonic is not included in the series expansion, since this
mode of deformation is already represented by re-scaling $R_{0}$ in equation
(\ref{rho}). The $l=1$ modes are also excluded, as $F$ will turn out to be
independent of them. These three modes describe a positive deformation on one
hemisphere of the vesicle, and negative on the opposite hemisphere. Hence, they
simply correspond to the three degrees of translational freedom. The complex
coefficients $\rho_{lm}$ become the dynamical variables, where
\begin{displaymath}
  \rho(\theta,\varphi)=\sum_{l=2}^{\infty} \sum_{m=-l}^l \rho_{lm}Y_{l}^m
  (\theta,\varphi).
\end{displaymath}
The deformation field $\rho$ is constrained to be real by demanding that
\begin{displaymath}
  \rho_{lm}^*=(-1)^m\rho_{l\,-m}
\end{displaymath}
which follows from the identity $Y_{l}^{m*}\equiv(-1)^m Y_{l}^{-m}$.

  As stated earlier, the order parameter field $\psi$ is expanded in the basis
functions given in equation (\ref{basis}) thus:
\begin{equation}
\label{expansion}
  \psi=\sqrt{\frac{4\pi}{\cal A}}\sum_{p=-n}^n a_{p}\phi_{p}
\end{equation}
where $a_{p}$ are $(2n+1)$ dimensionless, complex, dynamical variables.

  The free energy can now be expressed in terms of the dynamical variables
thus:
\begin{equation}
\label{free2}
  F=\alpha\sum_{p}a_{p}^*a_{p}+\sum_{pqrs}\beta_{pqrs}a_{p}^*a_{q}^*a_{r}a_{s}
  +\sum_{pqlm}\gamma_{pqlm}a_{p}^*a_{q}\rho_{lm}
  +\sum_{lm}\Delta_{l}\rho_{lm}^*\rho_{lm}
\end{equation}
where
\begin{mathletters}
\label{coeffs}
\begin{eqnarray}
  \alpha &=& r+\frac{4\pi n C}{\cal A}  \\
  \beta_{pqrs} &=& \frac{2\pi u}{\cal A}
  \int\phi_{p}^*\phi_{q}^*\phi_{r}\phi_{s}\,d\Omega   \\
  \gamma_{pqlm} &=& \frac{4\pi n C}{\cal A}(l+2)(l-1)
  \int\phi_{p}^*\phi_{q} Y_{l}^m \,d\Omega   \\
  \Delta_{l} &=& \mbox{$\frac{1}{2}$}\kappa l(l^2-1)(l+2)
\end{eqnarray}
\end{mathletters}
all of which coefficients are real.
Indices run over the intervals $-n\leq p,q,r,s\leq n$ and $2\leq l<\infty$
and $-l\leq m\leq l$. Notice that, as stated earlier, $l=1$ spherical harmonics
contribute to neither the curvature energy nor the coupling, due to the factors
of $(l-1)$. Notice also that equation (\ref{free2}) is quadratic in the
variables $\rho_{lm}$. Hence, in a partition function defined by
\begin{displaymath}
  {\cal Z} = \int e^{-F(\{a_{p}\},\{\rho_{lm}\})} \prod_{q\,l'm'}da_{q}^*da_{q}
  d\rho_{l'm'}^*d\rho_{l'm'}
\end{displaymath}
the integral over all shapes may be done explicitly by completing the square,
using the identity $\gamma_{pqlm}\equiv(-1)^m\gamma_{qpl-m}$.
Hence a new effective free energy, defined by
\begin{equation}
\label{partition}
  {\cal Z} = \int e^{-F^{\mbox{\tiny\it eff}}(\{a_{p}\})}
  \prod_{q}da_{q}^*da_{q}
\end{equation}
is found to be
\begin{equation}
\label{effective}
  F^{\mbox{\tiny\it eff}} =
  \alpha\sum_{p}a_{p}^*a_{p}+\sum_{pqrs}\beta^{\mbox{\tiny\it eff}}_{pqrs}
  a_{p}^*a_{q}^*a_{r}a_{s} + {\cal H}_{\rho}
\end{equation}
where
\begin{equation}
\label{beta}
  \beta^{\mbox{\tiny\it eff}}_{pqrs} = \beta_{pqrs}
  - \frac{1}{4}\sum_{l=2}^{\infty}
  \sum_{m=-l}^l \frac{\gamma_{rqlm}\gamma_{pslm}}{\Delta_{l}}
\end{equation}
and ${\cal H}_{\rho}$ is the free energy of shape fluctuations due to the
Helfrich Hamiltonian alone, given by
\begin{displaymath}
  {\cal H}_{\rho} =
  \sum_{l=2}^{\infty}\ln\left[\mbox{$\frac{1}{2}$}\left(\frac{\Delta_{l}}
  {\pi}\right)^{l+1}\right] = \zeta_{1}\ln\kappa + \zeta_{2}
\end{displaymath}
where $\zeta_{1}$ and $\zeta_{2}$ are infinite constants in the continuum
limit. This infinite part of the free energy will henceforth be renormalized
away.

  Consider for a moment the coefficients $\gamma_{pqlm}$, defined via the
integral $\int\phi_{p}^*\phi_{q}Y_{l}^md\Omega$. Now, the spherical harmonic
$Y_{l}^m(\theta,\varphi)$ is of the
form $P_{l}^m(cos\theta)\exp(im\varphi)$, where
$P_{l}^m$ is an associated Legendre function, of the form \mbox{$\sin^m\theta
\sum_{k=0}^{l-m}b_{k}\cos^k\theta$} with constant coefficients $b_{k}$. The
associated Legendre functions form a complete, orthogonal set of functions for
each value of $m$. Note also that
\mbox{$\phi_{p}^*\phi_{q}\propto\sin^{p-q}\theta (1-\cos\theta)^{n+q}
(1+\cos\theta)^{n-p}\exp i(q-p)\varphi$}. It
follows that it is possible to write
the function $\phi_{p}^*\phi_{q}$ as a {\em finite} sum of spherical harmonics.
{\em ie.} with some constant coefficients $c_{k}$, it can be written
\mbox{$\phi_{p}^*\phi_{q}=\sum_{j=0}^{2n}c_{j}Y_{j}^{p-q*}$}. Hence
$\phi_{p}^*\phi_{q}$ is orthogonal to all spherical harmonics not included in
this sum. So $\gamma_{pqlm}$ vanishes for $l>2n$, and the infinite sum in
equation (\ref{beta}) can be replaced by a finite sum
\begin{equation}
\label{betaeff}
  \beta^{\mbox{\tiny\it eff}}_{pqrs}
  = \beta_{pqrs} - \frac{1}{4}\sum_{l=2}^{2n}
  \sum_{m=-l}^l \frac{\gamma_{rqlm}\gamma_{pslm}}{\Delta_{l}} .
\end{equation}
Physically, this means that only the modes of deformation between $l=2$ and
$l=2n$ couple to the lowest Landau levels of order.

\section{Stability Analysis}

\label{stability}
  The original system of orientational order on a fluctuating sphere has now
been mapped onto a rigid sphere problem, in which $a_{p}$ are the only
dynamical variables, by replacing the coupling $\beta_{pqrs}$ by the effective
coupling $\beta^{\mbox{\tiny\it eff}}_{pqrs}$, renormalized by shape
fluctuations. The free energy closely resembles that of a superconducting film,
penetrated normally by magnetic flux quanta, around which the supercurrent
flows in vortices. It is well known \cite{Ruggeri76} that changing the sign of
the fourth-order term from positive to negative changes the nature of the
superconductor from type-II to type-I, from which all flux vortices are
excluded. The corresponding phenomenon of excluding all vortices from an
$n$-atic vesicle appears impossible, since the vortices are topologically
imperative. There is, in fact, an analogous exclusion of vortices, brought
about by a catastrophic change in the vesicle's topology, at which point the
model breaks down, since the order and deformation fields become large. In the
analysis of type-II superconductors, there is a similar break-down of the model
in the marginal limit. The criterion for a `type-I' vesicle is now found by a
stability analysis.

  The free energy of equation (\ref{effective}) is a function of the $(2n+1)$
complex variables $a_{p}$. It describes a stable system if $F$ is large at
infinity in all directions in the phase space, {\em ie.} if $F\rightarrow
+\infty$ as any linear combination of the variables $a_{p}$ goes to
infinity. Such a path to infinity may be parameterized
by $a_{p}=\eta_{p}\Xi$, where $\Xi\rightarrow\infty$. The constants
$\eta_{p}$ define a phase-space direction and may be kept
finite, without loss of generality, by the constraint
$\sum_{p}\eta_{p}^*\eta_{p}=1$. Substituting into equation
(\ref{effective}) and neglecting quantities of less than fourth order in $\Xi$
yields the criterion for stability
\begin{displaymath}
  \sum_{pqrs}\beta^{\mbox{\tiny\it eff}}_{pqrs}\eta_{p}^*\eta_{q}^*\eta_{r}
  \eta_{s}>0 \;\;\;\; \forall \{\eta_{p}\}.
\end{displaymath}
There is one particular set of constants $\eta_{p}=\eta^0_{p}$ which
minimizes this sum. This set definitely exists, due to the introduction of the
constraint above. So the stability criterion may be expressed in terms of this
specific set of constants thus:
\begin{displaymath}
  \sum_{pqrs}\beta^{\mbox{\tiny\it eff}}_{pqrs}\eta^{0*}_{p}\eta^{0*}_{q}
  \eta^0_{r}\eta^0_{s}>0.
\end{displaymath}
But this set of constants which minimizes the sum, subject to an overall
magnitude constraint, is simply the set of {\em mean-field} values of $a_{p}$,
multiplied by some scale factor. Hence the criterion for stability is
$\sum\beta^{\mbox{\tiny\it eff}}_{pqrs}(a^*_{p}a^*_{q}a_{r}a_{s}
)_{\mbox{\tiny M.F.}}>0$, and may be written as
\begin{displaymath}
  \int|\psi|^4_{\mbox{\tiny M.F.}} d\Omega
  >4\pi\mu\sum_{lm}\frac{(l+2)(l-1)}{l(l+1)}
  \left| \int|\psi|^2_{\mbox{\tiny M.F.}} Y_{l}^m d\Omega \right|^2.
\end{displaymath}
Let `malleability' $\mu$ be defined as the quantity $n^2C^2/{\cal A}u\kappa$,
which is a measure of the ease with which the order may deform the vesicle. The
marginal case, defining the crossover from `type-II' to `type-I' behaviour is
given by $\mu=\mu_{c}$ where
\begin{displaymath}
  \mu_{c}=\int|\psi|^4_{\mbox{\tiny M.F.}} d\Omega / 4\pi\sum_{lm}
  \frac{(l+2)(l-1)}{l(l+1)} \left| \int|\psi|^2_{\mbox{\tiny M.F.}}
  Y_{l}^m d\Omega \right|^2.
\end{displaymath} 
Using symmetry arguments to establish the mean-field configuration,
$\mu_{c}(n)$ is found for the first two cases to be: $\mu_{c}(1)=9/4\pi$,
$\mu_{c}(2)=125/1567\pi$. For vesicles
this malleable, the order parameter can deform the shape from a sphere to an
intrinsically flat cylinder or polyhedron, and hence ordering is no longer
frustrated by curvature. In other words, the vesicle bursts. This
interpretation may not be the whole story, since the model breaks down
as this instability is approached, due to the large magnitudes of the fields
$\psi$ and $\rho$. But it is clear that `ordinary' or `type-II' behaviour is
exhibited when $\mu<\mu_{c}$ and that a qualitatively different behaviour,
requiring a different model, exists in vesicles whose malleability is greater
than this finite threshold.
  It will be demonstrated later that ordering increases and the region of
temperature over which it arises sharpens with increasing malleability. The
relationship between the magnitude of this order and the degree of shape
deformation is now calculated.

\section{Shape Correlations}

  The original model for genus-zero vesicles with $n$-atic order, used by Park
et.\ al.\ \cite{Park92}, embodied in equations (\ref{F}) and (\ref{area}), has
been
reduced to the much simpler form of equation (\ref{simple})
and all of the shape
deformation degrees of freedom have been integrated out, leaving the effective
free energy in equation (\ref{effective}), with coefficients calculable from
equations (\ref{betaeff}) and (\ref{coeffs}).
Park et.\ al.\ solved the model in the zero temperature limit (without
fluctuations), and thus were able to produce diagrams of the vesicle shapes,
from the fixed values of $\rho_{lm}$ for each value of $n$. In reality of
course, no such fixed values exist, as the shapes and vortex positions are in a
constant state of flux, but expectation values of various moments of the
variables $\rho_{lm}$ and $a_{p}$ can be calculated. The variables $\rho_{lm}$
have been eliminated from the problem, so transformation equations must be
found, relating their expectation values to those of $a_{p}$. This is done with
the use of equation (\ref{free2}), and integrals of the form
\begin{displaymath}
  <\rho_{lm}>=\frac{1}{\cal Z}\int{\cal D}[\psi,\rho] \rho_{lm}e^{-F}
\end{displaymath}
with the results
\begin{equation}
\label{rho1}
  <\rho_{lm}> = \frac{-1}{2\Delta_{l}}\sum_{pq}\gamma_{pqlm}<a_{p}^*a_{q}>
\end{equation}
and
\begin{equation}
\label{rho2}
 <\rho_{l_{1}m_{1}}^*\rho_{l_{2}m_{2}}>=\frac{1}{4\Delta_{l_{1}}\Delta_{l_{2}}}
 \sum_{pqrs}\gamma_{qrl_{1}m_{1}}\gamma_{spl_{2}m_{2}}
 <a_{p}^*a_{q}^*a_{r}a_{s}>
 + \frac{\delta_{l_{1}l_{2}}\delta_{m_{1}m_{2}}}{\Delta_{l_{1}}}
\end{equation}
where $\delta_{xy}$ is the Kr\"{o}necker delta. The expression for the second
moment (equation (\ref{rho2})) is in two parts. The first represents
deformations
of the vesicle due to interaction with the order ({\em ie.} $\psi$ is
attempting to align, and expel Gaussian curvature, except at the vortices
where it is small). This part vanishes for higher spherical harmonics ($l>2n$),
which do not couple to the lowest landau levels.
The second part represents uncorrelated deformations due to thermal excitation.

\section{Validity of the Approximations}

\label{valid}
  Before proceeding further, an assessment is made of the regimes of validity
of the approximations employed so far.

  The Ginzburg-Landau model carries with it an implicit approximation. It
contains a truncated series expansion of the potential acting on $\psi$. For
this to be valid, it is required that higher order terms in the expansion are
smaller than lower order terms. This condition is satisfied at high temperature
where $\psi$ is small. It is also true at low temperature, provided that
\mbox{$(\mu_{c}-\mu)\tilde{>}\pi^{-1}$}. Clearly the model is good far from
`critical malleability', as stated above.

  In using the lowest-Landau-level approximation, it is assumed that the
`ground-state' modes dominate the order parameter field. Let the expectation
value of their coefficients be denoted $\left<a_{p_{0}}^*a_{p_{0}}\right>$ and
that of the next lowest level be denoted $\left<a_{p_{1}}^*a_{p_{1}}\right>$.
Then the ratio
$\left<a_{p_{1}}^*a_{p_{1}}\right>/\left<a_{p_{0}}^*a_{p_{0}}\right>$ must be
much less than unity. An order-of-magnitude estimate of this quantity is found
from a high-temperature calculation to be $\sim\alpha/(\alpha-r_{c})$, from
which it follows that $|\alpha|\ll C/{\cal A}$. So this approximation is valid
in the neighbourhood of the mean-field transition. Let its region of validity
cover a large domain in figures 1-4, so that all work presented here is
correct. This requires that $C^2/{\cal A}u\gg (\mu_{c}-\mu)$.

  One final approximation must be justified. In deriving equation
(\ref{simple}), a factor $(1+2\rho)$ was dropped from the line above, with the
explanation that $|\psi|^2\sim(r-r_{c})$, so this term was of too high an order
in small quantities. This is only apparent at low temperature, but the
derivation holds true at all temperatures for the following reasons. Replacing
this lost factor is equivalent to multiplying $\gamma_{pqlm}$ by
$1+2(1-\frac{r}{r_{c}})/(l+2)(l-1)$. Demanding that this is close to unity
leads to the condition $|\alpha|\ll C/{\cal A}$, which is identical to the
lowest-Landau-level criterion.

  It remains only to calculate expectation values of the order-parameter field
using equation (\ref{effective}). For $n=1$ (vector order) this is done in
section \ref{n=1} by making use of various symmetries of the system. For other
values of $n$, the calculation cannot be performed exactly, so a further
approximation will be introduced in section \ref{H-F}, where the problem is
solved for general $n$. Either of these sections may be read without reference
to the other.

\section{Non-Perturbative Solution for Vector Order}

\label{n=1}
  The $n=1$ case (Sm-C order), solved at mean-field level in
\cite{MacKintosh91}, has just two vortices on the spherical surface, and
therefore has more symmetries than higher-$n$ cases, which can be exploited in
the solution of the integrals of equation (\ref{partition}).

  A vesicle with vector order has two topologically required defects in the
order parameter field. It is modelled using three complex dynamical variables,
giving six degrees of freedom, corresponding to the positions of the two
vortices in the two-dimensional space, plus an overall complex amplitude. Its
energy is invariant under four obvious symmetries. The three Euler angles which
relate to the spatial orientation of the vesicle are clearly irrelevant degrees
of freedom, as is the complex phase of the overall amplitude of the order
parameter
field, which describes a global rotation of the orientation of in-plane order.
The two remaining relevant dynamical quantities are the real amplitude of the
order parameter and the geodesic distance between the vortices. A
transformation of variables is now performed on equations (\ref{partition}) and
(\ref{effective}) to make these symmetries explicit.

  Let us re-express the lowest-Landau-level order in the form used in
\cite{Park92}:
\begin{displaymath}
  \psi = \sqrt{\frac{12}{\cal A}} \psi_{0} \prod_{k=1}^{2} \left(
  \sin\frac{\theta}{2}\cos\frac{\theta_{k}}{2} e^{i(\varphi-\varphi_{k})/2}
 -\cos\frac{\theta}{2}\sin\frac{\theta_{k}}{2} e^{-i(\varphi-\varphi_{k})/2}
  \right)
\end{displaymath}
where $(\theta_{k},\varphi_{k})$ is the position of the $k^{\mbox{\tiny th}}$
vortex in spherical polar coordinates. The factor $\surd (12/{\cal A})$ is
convenient for the definition of the overall complex amplitude $\psi_{0}$.
Equating this expression to equation (\ref{expansion}) produces the
transformation equations:
\begin{mathletters}
\label{trans}
\begin{eqnarray}
  a_{1} &=& 2\psi_{0}\cos\frac{\theta_{1}}{2}\cos\frac{\theta_{2}}{2}
  e^{-i(\varphi_{1}+\varphi_{2})/2}   \\
  a_{-1} &=& 2\psi_{0}\sin\frac{\theta_{1}}{2}\sin\frac{\theta_{2}}{2}
  e^{i(\varphi_{1}+\varphi_{2})/2}    \\
  a_{0} &=& -\sqrt{2}\psi_{0}\left( \cos\frac{\theta_{1}}{2}
  \sin\frac{\theta_{2}}{2} e^{i(\varphi_{2}-\varphi_{1})/2} +
  \sin\frac{\theta_{1}}{2} \cos\frac{\theta_{2}}{2}
  e^{i(\varphi_{1}-\varphi_{2})/2} \right)
\end{eqnarray}
\end{mathletters}
which relate the six dynamical variables
$(a_{-1},a_{-1}^*,a_{0},a_{0}^*,a_{1},a_{1}^*)$ to the new variables
$(\psi_{0},\psi_{0}^*,\theta_{1},\theta_{2},\varphi_{1},\varphi_{2})$. These
transformation equations are now used to evaluate the terms of equation
(\ref{effective}). It is found that $\sum_{p}a_{p}^*a_{p}= |\psi_{0}|^2
(3+\cos\theta_{1}\cos\theta_{2} + \sin\theta_{1}\sin\theta_{2}
\cos(\varphi_{1}-\varphi_{2}))$. Let the positions of the vortices be expressed
in a coordinate-invariant manner, by defining a unit vector $\mbox{\boldmath
$n$}_{k}$ to point from the centre of the vesicle towards the $k^{\mbox{\tiny
th}}$ vortex. Then the cosine of the angle subtended by the vortices at the
centre is $(\mbox{\boldmath $n$}_{1}\cdot\mbox{\boldmath $n$}_{2})$. In
spherical polar coordinates, this becomes \mbox{$\mbox{\boldmath
$n$}_{1}\cdot\mbox{\boldmath $n$}_{2} = \cos\theta_{1}\cos\theta_{2} +
\sin\theta_{1}\sin\theta_{2} \cos(\varphi_{1}-\varphi_{2})$}. So this
coordinate-invariant description has arisen naturally from the model, and the
first term of equation (\ref{effective}) becomes
\begin{displaymath}
  \alpha\sum_{p=-1}^1a_{p}^*a_{p} = \alpha|\psi_{0}|^2
  (3+\mbox{\boldmath $n$}_{1}\cdot\mbox{\boldmath $n$}_{2}).
\end{displaymath}
Evaluation of the other terms of equation (\ref{effective}) is facilitated by
noting that
\begin{displaymath}
|\psi|^2=\frac{3}{\cal A}|\psi_{0}|^2 \prod_{k=1}^{2}
(1-\hat{\mbox{\boldmath $r$}}\cdot\mbox{\boldmath $n$}_{k})
\end{displaymath}
where {\boldmath $\hat{r}$} is the unit vector in the direction of
$\mbox{\boldmath $R$}$. Hence
\begin{eqnarray*}
  \sum_{pqrs}\beta^{\mbox{\tiny\it eff}}_{pqrs} a_{p}^*a_{q}^*a_{r}a_{s} &=& 
  \frac{{\cal A}u}{8\pi} \int|\psi|^4 d\Omega -\frac{C^2}{3\kappa}
  \sum_{m=-2}^2 \left| \int|\psi|^2 Y_{2}^m d\Omega \right|^2    \\
  &=& \frac{u|\psi_{0}|^4}{15{\cal A}} \left\{
  9 \left( 13+10\mbox{\boldmath $n$}_{1}\cdot\mbox{\boldmath $n$}_{2}
  +(\mbox{\boldmath $n$}_{1}\cdot\mbox{\boldmath $n$}_{2})^2 \right)
  - 4\pi\mu \left( 3+(\mbox{\boldmath $n$}_{1}\cdot\mbox{\boldmath $n$}_{2})^2
  \right) \right\}.
\end{eqnarray*}
As one would hope, all parts of the free energy are expressible in a
coordinate-invariant way. To complete the transformation of variables in
equation (\ref{partition}), a Jacobian must be evaluated. By calculation of a
six-by-six determinant, it is established from the transformation equations
(\ref{trans}) that
\begin{equation}
\label{Jacobian}
  \left| \frac{\partial(a_{-1}^*,a_{-1},a_{0}^*,a_{0},a_{1}^*,a_{1})}
  {\partial(\psi_{0}^*,\psi_{0},\theta_{1},\theta_{2},\varphi_{1},\varphi_{2})}
  \right| = |\psi_{0}|^4
  (1-\mbox{\boldmath $n$}_{1}\cdot\mbox{\boldmath $n$}_{2})\sin\theta_{1}
  \sin\theta_{2}.
\end{equation}
Finally, let $\psi_{0}=t e^{i\sigma}$ so that $t$ is a
measure of the overall real amplitude of the order parameter.
The partition function for smectic-C vesicles becomes
\begin{displaymath}
  {\cal Z} = 2\pi \int
  e^{-F^{\mbox{\tiny\it eff}}(t,\mbox{\boldmath $n$}_{1}\cdot
  \mbox{\boldmath $n$}_{2})} t^5
  (1-\mbox{\boldmath $n$}_{1}\cdot\mbox{\boldmath $n$}_{2}) \, d\Omega_{1}
  d\Omega_{2} dt
\end{displaymath}
where $d\Omega_{k}=\sin\theta_{k} d\theta_{k}d\varphi_{k}$.
As noted above, the free energy is independent of the overall complex phase
$\sigma$, which has been integrated out of the partition function, giving rise
to the factor $2\pi$. Integrals over both vortex positions still remain,
although it is only their {\em relative} separation that is relevant. Let
$\chi$ be a measure of this relative separation, being equal to the cosine of
the angle subtended by the vortices at the centre of the vesicle. Hence $\chi$
ranges in value from -1 when the vortices are antipodal to +1 when they are
coincident. This variable is conveniently introduced into the partition
function thus:
\begin{displaymath}
  {\cal Z} = 2\pi \int
  e^{-F^{\mbox{\tiny\it eff}}(t,\mbox{\boldmath $n$}_{1}\cdot
  \mbox{\boldmath $n$}_{2})} t^5
  (1-\mbox{\boldmath $n$}_{1}\cdot\mbox{\boldmath $n$}_{2}) \, dt \,
  d\Omega_{1} d\Omega_{2}\,
  \delta(\mbox{\boldmath $n$}_{1}\cdot\mbox{\boldmath $n$}_{2}-\chi) d\chi.
\end{displaymath}
Hence
\begin{equation}
\label{Z}
  {\cal Z} = 16\pi^3 \int_{-1}^1 \!\!\!\! d\chi \,
  \int_{0}^{\infty} \!\!\!\! dt \; t^5 (1-\chi)\,
  e^{-F^{\text{\it eff}}(t,\chi)} 
\end{equation}
\mbox{where $F^{\text{\it eff}}=A(\chi)t^2+B(\chi)t^4$ with $A=\alpha(3+\chi)$
and}
$B=\left\{9(13+10\chi+\chi^2)-4\pi\mu(3+\chi^2)\right\}/15\omega^2$. The
original set of
six thermodynamic parameters $\{{\cal A},r,u,C,\kappa\}$ has been reduced to
just three independent combinations:
\begin{itemize}
  \item the shifted temperature-like parameter \mbox{$\alpha\equiv r+4\pi
nC/{\cal A}$},
  \item the malleability $\mu\equiv n^2C^2/{\cal A}u\kappa$,
  \item and the scale parameter $\omega\equiv\sqrt{{\cal A}/u}$,
\end{itemize}
with $n=1$ in this case.

  Notice that there is more phase space available to
vortices at large separations than to vortices in close proximity; a phenomenon
arising from the phase space factor $(1-\chi)$. One might na\"{\i}vely predict
that the thermal expectation value of $\chi$ would tend to zero in the limit of
high temperature since the vortices would spend as much time close together as
they would in opposite hemispheres. But this is not the case. The factor
$(1-\chi)$ favours large inter-vortex distances, and
hence $\left<\chi\right>$ must tend
to a finite, negative number at high temperature.

  The double integral of equation (\ref{Z})
is soluble; the method being given in
appendix \ref{integral1}. The solution is
\begin{equation}
\label{solution}
  {\cal Z} = \frac{10\pi^{\frac{7}{2}}\omega^2}{(3+2\pi\mu)\alpha}
  \left[ f\!\left( \frac{\alpha\omega}{\sqrt{\frac{4}{15}(9-4\pi\mu)}} \right)-
  f\!\left( \frac{\alpha\omega}{\sqrt{\frac{2}{15}(27-2\pi\mu)}} \right)\right]
\end{equation}
where $f(x) \equiv x\,e^{x^2}\!\mbox{\,erfc\,}{x}\:$ and $\mbox{\,erfc\,}$ is
the complementary error function: $\mbox{\,erfc\,}{x}=1-\mbox{\,erf\,}{x}$. As
$x\rightarrow+\infty$, $f(x)\rightarrow(1-1/2x^2)/\sqrt{\pi}$ and as
$x\rightarrow-\infty$, $f(x)\rightarrow 2x\,e^{x^2}$.

  From the expressions for the partition function in equations
(\ref{partition})
and (\ref{Z}), the following relations can be deduced:
\begin{mathletters}
\label{derivatives}
\begin{eqnarray}
  -\frac{\partial\ln{\cal Z}}{\partial\alpha} &=& \left<(3+\chi)t^2\right> =
  \left<\sum_{p}a_{p}^*a_{p}\right> = \left<\int|\psi|^2 d{\cal A}\right> \\
  \frac{5\omega^3}{6}\frac{\partial\ln{\cal Z}}{\partial\omega} &=&
  \left<(13+10\chi+\chi^2)t^4\right> \\
  \frac{15\omega^2}{4\pi}\frac{\partial\ln{\cal Z}}{\partial\mu} &=&
  \left<(3+\chi^2)t^4\right>
\end{eqnarray}
\end{mathletters}
from which various correlators of the variables $a_{p}$ are derived below.
However, moments of $\chi$ cannot be found from ${\cal Z}$ alone and, as noted
in appendix \ref{integral1}, it is not easy to simply invent a field coupled to
$\chi$ to make this possible.

  Expectation values of powers of $\chi$ are evaluated as follows. Returning to
equation (\ref{Z}), the $t$ integral alone is easily solved, giving
\begin{equation}
\label{Z1}
  {\cal Z} = 2\pi^{\frac{7}{2}}\int_{-1}^1 (1-\chi) B^{-\frac{3}{2}}
  f'\left(\frac{A}{2\sqrt{B}}\right)\, d\chi
\end{equation}
where $f'(x)$ is the derivative of the function $f(x)$ defined above. Hence the
moments of $\chi$ are given by
\begin{displaymath}
  \left<\chi^m\right>= \frac{2\pi^{\frac{7}{2}}}{\cal Z}
  \int_{-1}^1 \chi^m (1-\chi) B^{-\frac{3}{2}}
  f'\left(\frac{A}{2\sqrt{B}}\right)\, d\chi
\end{displaymath}
which is a function of $\mu$ and $(\alpha\omega)$ only. A graph of
$\left<\chi\right>$ against $\alpha\omega$ is plotted in figure \ref{chi} for
the cases of the rigid sphere ($\mu=0$), the marginal type-I-type-II vesicle
($\mu=\mu_{c}$), and an intermediate malleability ($\mu=\mu_{c}/2$). As $\mu$
increases, the vesicle becomes less rigid, and the lower `kink' in the graph
becomes sharper until, as $\mu\rightarrow 9/4\pi$, $\left<\chi\right>$ tends to
a singular function for which the mean-field value $\left<\chi\right>=-1$ is
correct for negative $\alpha$, as the above integrands are then singular for
this value of $\chi$. In fact, it is generally true that mean field theory
becomes increasingly accurate for negative $\alpha$ as $\mu$ approaches this
critical value. The value $\mu_{c}=9/4\pi$ is in agreement with the critical
malleability calculated in section \ref{stability}. For all $\mu$, at high
temperature ({\em ie.\ }in the large $\alpha\omega$ limit), $\left<\chi\right>$
tends to \mbox{$5-8\ln2\approx{-0.545}$}, which is a finite, negative number by
virtue of the phase space factor discussed above. Consider equation (\ref{Z1})
once more. Writing ${\cal Z}=\int d\chi\exp{-V(\chi)}$, we see that the
effective inter-vortex potential,
\begin{displaymath}
  V(\chi)=-\ln f'\!\left(\frac{A(\chi)}{2\sqrt{B(\chi)}}\right)
  +\mbox{$\frac{3}{2}$}\ln B - \ln(1-\chi)
\end{displaymath}
is repulsive. For this reason, $\left<\chi\right>\rightarrow{-1}$ in the
low-temperature limit, as this is the antipodal configuration. Note that
$V(\chi)$ is negatively dependent on $\mu$. So the vortex-vortex repulsion is
weakened when it is renormalized by shape fluctuations.

  Let us return to the problem of calculating expectation values from the
partition function as expressed in equation (\ref{solution}). Equations
(\ref{derivatives}) may be used to find a certain set of expectation values,
but
these are not in the form required by equations (\ref{rho1}) and (\ref{rho2})
for
evaluating expectation values of the vesicle's shape. How are the correlators
$\left<a^*_{p}a_{q}\right>$ and $\left<a^*_{p}a^*_{q}a_{r}a_{s}\right>$ to be
deduced from the combinations of $\chi$ and $t$ produced by equations
(\ref{derivatives})? One can quickly convince oneself, from the form of
equation (\ref{effective}),
with its `momentum' conserving coupling, that $\left<a^*_{p}a_{q}\right>=0$ for
$p\neq q$. Let ${\cal G}_{p}(t,\chi)$ be the mean value of $a^*_{p}a_{p}$, when
thermally averaged over all configurations which have a particular value of $t$
and $\chi$. This constrained average is given by
\begin{displaymath}
  {\cal G}_{p}(t_{0},\chi_{0}) = \frac{\int a_{p}^*a_{p}\,
  e^{-F^{\mbox{\tiny\it eff}}}
  \,\delta(\chi-\chi_{0}) \,\delta(t-t_{0}) \prod_{q} da_{q}^* da_{q}}
  {\int e^{-F^{\mbox{\tiny\it eff}}} \,\delta(\chi-\chi_{0}) \,\delta(t-t_{0})
  \prod_{q} da_{q}^* da_{q}}.
\end{displaymath}
Using the Jacobian of equation (\ref{Jacobian}) to change the variables of
integration, the constrained average is found to be
\begin{displaymath}
  {\cal G}_{p}(t_{0},\chi_{0}) = \frac{1}{8\pi^2}\int a_{p}^* a_{p}
  \,\delta(\mbox{\boldmath $n$}_{1}\cdot\mbox{\boldmath $n$}_{2}-\chi_{0})
  d\Omega_{1} d\Omega_{2}.
\end{displaymath}
And similarly, defining ${\cal G}_{pq}(t,\chi)$ to be given by thermally
averaging $a_{p}^*a_{p}a_{q}^*a_{q}$ over all states of a given $t$ and $\chi$,
it is found that
\begin{displaymath}
  {\cal G}_{pq}(t_{0},\chi_{0}) = \frac{1}{8\pi^2}\int a_{p}^*a_{p}a_{q}^*a_{q}
  \,\delta(\mbox{\boldmath $n$}_{1}\cdot\mbox{\boldmath $n$}_{2}-\chi_{0})
  d\Omega_{1} d\Omega_{2}
\end{displaymath}
all other fourth-order moments being zero. Having found these constrained
averages, the full thermal
averages of $a_{p}^*a_{p}$ and $a_{p}^*a_{p}a_{q}^*a_{q}$ will result from
averaging the functions ${\cal G}_{p}(t,\chi)$ and ${\cal G}_{pq}(t,\chi)$ over
$t$ and $\chi$. Hence, this is a derivation of transformation equations between
correlators of $a_{p}$ and correlators of $t$ and $\chi$.
From equations (\ref{derivatives}),
\begin{eqnarray*}
  a_{1}^*a_{1} &=& t^2 (1+\cos\theta_{1})(1+\cos\theta_{2}) \\
  a_{-1}^*a_{-1} &=& t^2 (1-\cos\theta_{1})(1-\cos\theta_{2}) \\
  a_{0}^*a_{0} &=& t^2 (1-2\cos\theta_{1}\cos\theta_{2}
  + \mbox{\boldmath $n$}_{1}\cdot\mbox{\boldmath $n$}_{2}).
\end{eqnarray*}
So, in general, the integrals required are of the form \mbox{$\int
g(\cos\theta_{1},\cos\theta_{2}) \,\delta(\mbox{\boldmath
$n$}_{1}\cdot\mbox{\boldmath $n$}_{2}-\chi) d\Omega_{1} d\Omega_{2}$} where
$g(x,y)$ is some function. This integral then, is of a function of the
positions of two points on a unit sphere. The points are each varied over the
surface, in such a way that their separation remains constant. A general
solution for such integrals is calculated in appendix \ref{integral2}. The
solution is
\begin{displaymath}
  \int g(\cos\theta_{1},\cos\theta_{2})
  \,\delta(\mbox{\boldmath $n$}_{1}\cdot\mbox{\boldmath $n$}_{2}-\chi)
  d\Omega_{1} d\Omega_{2} = 4\pi\int_{0}^{1}\!\!\!ds \int_{0}^{2\pi}\!\!\!d\xi
  \frac{s}{\sqrt{1-s^2}}g(x,y) 
\end{displaymath}
where
\begin{eqnarray*}
  x &=& \mbox{$\frac{1}{2}$} s \left[ (\sqrt{1+\chi}+\sqrt{1-\chi})\cos\xi
  + (\sqrt{1+\chi}-\sqrt{1-\chi})\sin\xi \right] \\
  y &=& \mbox{$\frac{1}{2}$} s \left[ (\sqrt{1+\chi}-\sqrt{1-\chi})\cos\xi
  + (\sqrt{1+\chi}+\sqrt{1-\chi})\sin\xi \right].
\end{eqnarray*}
Finally, the following correlation functions are found:
\begin{displaymath}
  \left<a_{p}^*a_{p}\right>=\left<t^2\right>
  +\mbox{$\frac{1}{3}$}\left<t^2\chi\right>
  = -\mbox{$\frac{1}{3}$}\frac{\partial\ln{\cal Z}}{\partial\alpha}
  \;\;\;\;\forall p
\end{displaymath}
and
\begin{eqnarray*}
  \left<a_{-1}^*a_{-1}a_{-1}^*a_{-1}\right> =
  \left<a_{1}^*a_{1}a_{1}^*a_{1}\right>
  &=& 2\left<a_{-1}^*a_{-1}a_{0}^*a_{0}\right> \\
  = 2\left<a_{1}^*a_{1}a_{0}^*a_{0}\right>
  &=& \mbox{$\frac{2}{15}$}\left( 13\left<t^4\right>+10\left<t^4\chi\right>
  +\left<t^4\chi^2\right>\right)
  = \frac{\omega^3}{9}\frac{\partial\ln{\cal Z}}{\partial\omega}      \\
  \left<a_{-1}^*a_{-1}a_{1}^*a_{1}\right> &=& \mbox{$\frac{2}{15}$}
  \left( 3\left<t^4\right>+\left<t^4\chi^2\right>
  \right) = \frac{\omega^2}{2\pi}\frac{\partial\ln{\cal Z}}{\partial\mu} \\
  \left<a_{0}^*a_{0}a_{0}^*a_{0}\right>
  &=& \left<a_{-1}^*a_{-1}a_{1}^*a_{1}\right>
  +\mbox{$\frac{1}{2}$}\left<a_{1}^*a_{1}a_{1}^*a_{1}\right>.
\end{eqnarray*}
  The derivatives are performed on equation (\ref{solution}).
The order parameter is found to behave as follows:
\begin{eqnarray*}
  & & \frac{1}{\omega}\left<\int\!\!|\psi|^2\,d{\cal A}\right> =
  \frac{1}{\alpha\omega} \\
  & & - \left(\frac{f'\left(\frac{\alpha\omega}
  {\sqrt{\frac{4}{15}(9-4\pi\mu)}}\right)}
  {\sqrt{\frac{4}{15}(9-4\pi\mu)}}
  - \frac{f'\left(\frac{\alpha\omega}{\sqrt{\frac{2}{15}(27-2\pi\mu)}}\right)}
  {\sqrt{\frac{2}{15}(27-2\pi\mu)}}\right) / \left(
  f\left(\frac{\alpha\omega}{\sqrt{\frac{4}{15}(9-4\pi\mu)}}\right)
  -f\left(\frac{\alpha\omega}{\sqrt{\frac{2}{15}(27-2\pi\mu)}}\right)\right).
\end{eqnarray*}
In the high-temperature limit, this tends to $3/\alpha\omega$ and, in the
low-temperature limit, to the mean-field value of
$-15\alpha\omega/(18-8\pi\mu)$. So, as $\mu$ increases towards the critical
value of $\mu_{c}=9/4\pi$, the order parameter increases in magnitude at low
temperature, but not at high temperature, so the crossover between the two
regimes becomes sharper, and mean-field theory becomes relatively more
accurate. Figure \ref{order1} is a graph of the quantity
\mbox{$\left<\int|\psi^2|d{\cal A}\right>(18-8\pi\mu)/15\omega$}, which is
normalized for low temperature, against $\alpha\omega$, for various values of
the malleability $\mu$.

  The shape expectation values can now be found. As all of the correlators
$\left<a_{p}^*a_{p}\right>$ are equal, substituting for
$\left<a_{p}^*a_{q}\right>$ in
equation (\ref{rho1}) with $\left<a_{p}^*a_{p}\right>\delta_{pq}$ gives
\begin{displaymath}
  \left<\rho_{lm}\right>=\frac{-3\left<t^2\right>
  -\left<t^2\chi\right>}{6\Delta_{l}}\sum_{p}\gamma_{pplm}.
\end{displaymath}
But $\sum_{p}\gamma_{pplm}=0$. So all spherical harmonic amplitudes have a
mean value of zero, despite the distorting influence of the two topological
defects in $\psi$. This is because all vortex positions have been thermally
averaged. The second moments of $\rho$ are calculable, via equation
(\ref{rho2}),
from the fourth moments of the order parameter field. For instance, the
mean-square amplitude of the $Y_{2}^0$ deformation mode is found to obey
\begin{eqnarray*}
  \kappa\left<|\rho_{20}|^2\right> &=& \frac{1}{12}  \\
  &+& \frac{\pi}{45}\mu\alpha\omega
  \left( \frac{ 4 [\frac{4}{15}(9-4\pi\mu)]^{-3/2}
  f'\!\left( \frac{\alpha\omega}{\sqrt{\frac{4}{15}(9-4\pi\mu)}} \right)
  - [\frac{2}{15}(27-2\pi\mu)]^{-3/2}
  f'\!\left( \frac{\alpha\omega}{\sqrt{\frac{2}{15}(27-2\pi\mu)}} \right) }
  { f\!\left( \frac{\alpha\omega}{\sqrt{\frac{4}{15}(9-4\pi\mu)}} \right)
  - f\!\left( \frac{\alpha\omega}{\sqrt{\frac{2}{15}(27-2\pi\mu)}} \right) }
  \right).
\end{eqnarray*}
The constant $\frac{1}{12}$ is due to thermal excitation, and the other term
arises from deformation by the in-plane order. As $\mu\rightarrow\mu_{c}$, this
function goes to zero for positive $\alpha$, and to
$\frac{45}{128\pi^2}\left(\frac{\alpha\omega}{\mu_{c}-\mu}\right)^2$ for
negative $\alpha$. The latter expression tends to infinity as expected, since
deformations become large as the marginal type-I-type-II case is approached.
The function $(\mu_{c}-\mu)^2\kappa\left<|\rho_{20}|^2\right>$, which
is normalized for low temperature, is plotted against $\alpha\omega$ in figure
\ref{deform} for the cases $\mu=\mu_{c}/2$ and $\mu=\mu_{c}$.

  Notice that all of the measurable quantities calculated in this section are
functions of $\mu$ and the combination $\alpha\omega$. Hence, temperature,
measured from the mean-field transition, scales as $\omega^{-1}$. So
`transitions' from low- to high-temperature behaviour become sharper for larger
vesicles ({\em ie.} when ${\cal A}$ is large compared with $u$).

\section{Hartree-Fock Approximation}

\label{H-F}
  Up to this point, the only
approximation employed, other than the small amplitude approximations implicit
in the Ginzburg-Landau model, has been the confinement of the order parameter
field $\psi$ to a $2(2n+1)$-dimensional phase space of the lowest
eigenfunctions of the gradient operator.  A further approximation is now
introduced, in order to find expectation values of the $n$-atic order-parameter
field for all of the values of $n$ under consideration.
The Hartree-Fock method will be used to produce an infinite, but
incomplete, perturbation expansion, valid at high temperature - the opposite
limit to that studied in \cite{Park92}.

  Firstly, the bare, or Gaussian propagator $h$ is calculated from the
quadratic Hamiltonian
\begin{displaymath}
  {\cal H}_{0} = \alpha\sum_{p}a_{p}^*a_{p}
\end{displaymath}
and is found to be independent of $p$.
\begin{displaymath}
  \left<a_{p}^*a_{q}\right>_{0}\equiv
  \frac{\int a_{p}^*a_{q}e^{-{\cal H}_{0}}\prod_{r}da_{r}^*da_{r}}
  {\int e^{-{\cal H}_{0}}\prod_{r}da_{r}^*da_{r}} = \frac{\delta_{pq}}{\alpha}
  \equiv \delta_{pq}h_{p}
\end{displaymath}
Writing the free energy as $F={\cal H}_{0}+{\cal H}_{1}$ where ${\cal
H}_{1}=\sum_{pqrs}\beta^{\mbox{\tiny\it eff}}a_{p}^*a_{q}^*a_{r}a_{s}$, the
partition function becomes
\begin{displaymath}
  {\cal Z} = {\cal Z}_{0}\left<e^{-{\cal H}_{1}}\right>_{0}
\end{displaymath}
where ${\cal Z}_{0}$ is the partition function for the quadratic Hamiltonian
and $\left<\right>_{0}$ indicates a thermal average with respect to this
Hamiltonian. Hence, the renormalized propagator ({\em ie.} the full thermal
average, can be written as
\begin{displaymath}
  \tilde{h}_{p}\equiv\left<a_{p}^*a_{p}\right>=
  \frac{\left<a_{p}^*a_{p}e^{-{\cal H}_{1}}\right>_{0}}
  {\left<e^{-{\cal H}_{1}}\right>_{0}}.
\end{displaymath}
Taylor-expanding the exponentials and applying Wick's theorem, the terms in
this formula may be represented by connected, two-leg graphs of all possible
topologies, in which directed lines represent the bare propagators with
`momentum' $p$, and the four-point vertices each carry a factor of
$\beta^{\mbox{\tiny\it eff}}$. Momentum is conserved at the vertices, since
$\beta^{\mbox{\tiny\it eff}}_{pqrs}$ vanishes if \mbox{$p+q\neq r+s$.} The
numerator is the sum of all connected and disconnected two-leg graphs, but the
disconnected parts cancel with the denominator, which is the sum of all
zero-leg graphs. This explanation is brief, since the Feynman-diagram expansion
is given in many standard texts \cite{Ma}.

  Truncating the expansion at one-loop graphs yields
\begin{eqnarray*}
  \tilde{h}_{t} &\approx& h_{t} - \sum_{pqrs}^{\mbox{\tiny connected}}
   \beta_{pqrs}\left<a_{t}^*a_{p}^*a_{q}^*a_{r}a_{s}a_{t}\right>_{0} \\
  &=& h_{t} - \sum_{pqrs}^{\mbox{\tiny connected}} \beta_{pqrs} h_{t}h_{p}h_{q}
   (\delta_{tr}+\delta_{ts})(\delta_{pr}\delta_{qs}+\delta_{ps}\delta_{qr})
\end{eqnarray*}
represented by

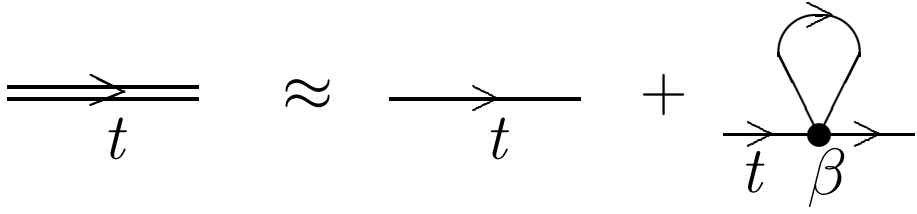
\setlength{\unitlength}{0.012500in}%
\begingroup\makeatletter\ifx\SetFigFont\undefined
\def\x#1#2#3#4#5#6#7\relax{\def\x{#1#2#3#4#5#6}}%
\expandafter\x\fmtname xxxxxx\relax \def\y{splain}%
\ifx\x\y   
\gdef\SetFigFont#1#2#3{%
  \ifnum #1<17\tiny\else \ifnum #1<20\small\else
  \ifnum #1<24\normalsize\else \ifnum #1<29\large\else
  \ifnum #1<34\Large\else \ifnum #1<41\LARGE\else
     \huge\fi\fi\fi\fi\fi\fi
  \csname #3\endcsname}%
\else
\gdef\SetFigFont#1#2#3{\begingroup
  \count@#1\relax \ifnum 25<\count@\count@25\fi
  \def\x{\endgroup\@setsize\SetFigFont{#2pt}}%
  \expandafter\x
    \csname \romannumeral\the\count@ pt\expandafter\endcsname
    \csname @\romannumeral\the\count@ pt\endcsname
  \csname #3\endcsname}%
\fi
\fi\endgroup
\begin{picture}(380,90)(10,590)
\thinlines
\put(380,620){\circle*{10}}
\thicklines
\put( 40,640){\line( 1, 0){ 80}}
\put( 75,645){\line( 2,-1){ 14}}
\put( 89,638){\line(-5,-2){ 14.138}}
\put( 40,635){\line( 1, 0){ 80}}
\put(200,635){\line( 1, 0){ 80}}
\multiput(235,640)(0.50000,-0.25000){21}{\makebox(0.4444,0.6667)
{\SetFigFont{7}{8.4}{rm}.}}
\multiput(245,635)(-0.50000,-0.25000){21}{\makebox(0.4444,0.6667)
{\SetFigFont{7}{8.4}{rm}.}}
\put(340,620){\line( 1, 0){ 80}}
\multiput(350,625)(0.50000,-0.25000){21}{\makebox(0.4444,0.6667)
{\SetFigFont{7}{8.4}{rm}.}}
\multiput(360,620)(-0.50000,-0.25000){21}{\makebox(0.4444,0.6667)
{\SetFigFont{7}{8.4}{rm}.}}
\put(380,653){\oval( 34, 34)[tr]}
\put(380,653){\oval( 34, 34)[tl]}
\multiput(395,625)(0.50000,-0.25000){21}{\makebox(0.4444,0.6667)
{\SetFigFont{7}{8.4}{rm}.}}
\multiput(405,620)(-0.50000,-0.25000){21}{\makebox(0.4444,0.6667)
{\SetFigFont{7}{8.4}{rm}.}}
\put(375,595){\makebox(0,0)[lb]{\smash{\SetFigFont{29}{34.8}{rm}$\beta$}}}
\put(380,620){\line(-1, 2){ 16.7}}
\multiput(375,675)(0.50000,-0.25000){21}{\makebox(0.4444,0.6667)
{\SetFigFont{7}{8.4}{rm}.}}
\multiput(385,670)(-0.50000,-0.25000){21}{\makebox(0.4444,0.6667)
{\SetFigFont{7}{8.4}{rm}.}}
\put(380,620){\line( 1, 2){ 16.7}}
\put(305,630){\makebox(0,0)[lb]{\smash{\SetFigFont{29}{34.8}{rm}+}}}
\put(155,630){\makebox(0,0)[lb]{\smash{\SetFigFont{29}{34.8}{rm}$\approx$}}}
\put(347,596){\makebox(0,0)[lb]{\smash{\SetFigFont{29}{34.8}{it}t}}}
\put(240,610){\makebox(0,0)[lb]{\smash{\SetFigFont{29}{34.8}{it}t}}}
\put( 80,610){\makebox(0,0)[lb]{\smash{\SetFigFont{29}{34.8}{it}t}}}
\end{picture}

\begin{flushleft}
where a double line represents a renormalized propagator.
\end{flushleft}

  The Hartree-Fock scheme renormalizes the bare propagator with not just one
loop diagram, but an infinity of diagrams belonging to a certain topological
family. In particular, all those diagrams containing loops which do not span a
vertex. This approximation to the renormalized propagator is defined implicitly
by the equation

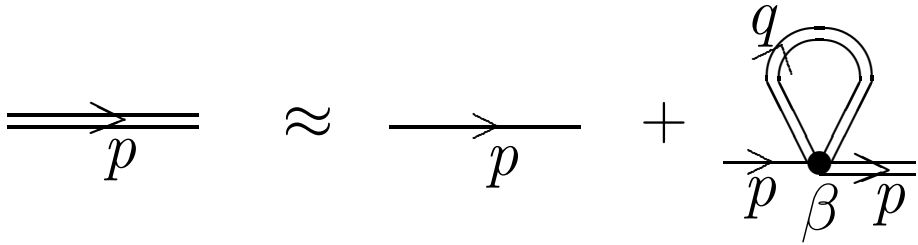
\setlength{\unitlength}{0.012500in}%
\begingroup\makeatletter\ifx\SetFigFont\undefined
\def\x#1#2#3#4#5#6#7\relax{\def\x{#1#2#3#4#5#6}}%
\expandafter\x\fmtname xxxxxx\relax \def\y{splain}%
\ifx\x\y   
\gdef\SetFigFont#1#2#3{%
  \ifnum #1<17\tiny\else \ifnum #1<20\small\else
  \ifnum #1<24\normalsize\else \ifnum #1<29\large\else
  \ifnum #1<34\Large\else \ifnum #1<41\LARGE\else
     \huge\fi\fi\fi\fi\fi\fi
  \csname #3\endcsname}%
\else
\gdef\SetFigFont#1#2#3{\begingroup
  \count@#1\relax \ifnum 25<\count@\count@25\fi
  \def\x{\endgroup\@setsize\SetFigFont{#2pt}}%
  \expandafter\x
    \csname \romannumeral\the\count@ pt\expandafter\endcsname
    \csname @\romannumeral\the\count@ pt\endcsname
  \csname #3\endcsname}%
\fi
\fi\endgroup
\begin{picture}(380,108)(40,595)
\thinlines
\put(380,620){\circle*{10}}
\thicklines
\put(200,635){\line( 1, 0){ 80}}
\put( 40,635){\line( 1, 0){ 80}}
\put( 40,640){\line( 1, 0){ 80}}
\put( 75,645){\line( 2,-1){ 14}}
\put( 89,638){\line(-5,-2){ 14.138}}
\put(380,615){\line( 1, 0){ 40}}
\put(380,620){\line( 1, 0){ 40}}
\put(340,620){\line( 1, 0){ 40}}
\put(395,624){\line( 2,-1){ 14}}
\put(409,617){\line(-5,-2){ 14.138}}
\put(380,654.5){\oval( 44, 44)[tr]}
\put(380,654.5){\oval( 44, 44)[tl]}
\put(380,654.5){\oval( 34, 34)[tr]}
\put(380,654.5){\oval( 34, 34)[tl]}
\put(375,620){\line(-1, 2){ 17}}
\put(380,620){\line(-1, 2){ 17}}
\put(380,620){\line( 1, 2){ 17}}
\put(385,620){\line( 1, 2){ 17}}
\multiput(352,663)(0.50000,0.25000){25}{\makebox(0.4444,0.6667)
{\SetFigFont{7}{8.4}{rm}.}}
\multiput(364,669)(0.18182,-0.54545){23}{\makebox(0.4444,0.6667)
{\SetFigFont{7}{8.4}{rm}.}}
\multiput(350,625)(0.50000,-0.25000){21}{\makebox(0.4444,0.6667)
{\SetFigFont{7}{8.4}{rm}.}}
\multiput(360,620)(-0.50000,-0.25000){21}{\makebox(0.4444,0.6667)
{\SetFigFont{7}{8.4}{rm}.}}
\multiput(235,640)(0.50000,-0.25000){21}{\makebox(0.4444,0.6667)
{\SetFigFont{7}{8.4}{rm}.}}
\multiput(245,635)(-0.50000,-0.25000){21}{\makebox(0.4444,0.6667)
{\SetFigFont{7}{8.4}{rm}.}}
\put(155,630){\makebox(0,0)[lb]{\smash{\SetFigFont{29}{34.8}{rm}$\approx$}}}
\put(305,630){\makebox(0,0)[lb]{\smash{\SetFigFont{29}{34.8}{rm}+}}}
\put(372,592){\makebox(0,0)[lb]{\smash{\SetFigFont{29}{34.8}{it}$\beta$}}}
\put( 80,618){\makebox(0,0)[lb]{\smash{\SetFigFont{29}{34.8}{it}p}}}
\put(240,615){\makebox(0,0)[lb]{\smash{\SetFigFont{29}{34.8}{it}p}}}
\put(348,602){\makebox(0,0)[lb]{\smash{\SetFigFont{29}{34.8}{it}p}}}
\put(402,599){\makebox(0,0)[lb]{\smash{\SetFigFont{29}{34.8}{it}p}}}
\put(349,674){\makebox(0,0)[lb]{\smash{\SetFigFont{29}{34.8}{it}q}}}
\end{picture}

\begin{flushleft}
which can be thought of as iteratively replacing every bare propagator with the
one-loop correction. The scheme generates graphs with the correct counting
factors \cite{Ruggeri76}. This pictorial equation corresponds to
\end{flushleft}
\begin{displaymath}
  \tilde{h}_{p}\approx h_{p}\left(1-2\tilde{h}_{p}\sum_{q}
  (\beta^{\mbox{\tiny\it eff}}_{pqqp}
  +\beta^{\mbox{\tiny\it eff}}_{pqpq})\tilde{h}_{q}\right)
\end{displaymath}
a set of $(2n+1)$ second order coupled polynomial equations. These generate
many spurious solutions. But it is observed, from calculation of
$\beta_{pqrs}^{\mbox{\tiny\it eff}}$ that
$\sum_{q}(\beta_{pqqp}^{\mbox{\tiny\it eff}}+\beta_{pqpq}^{\mbox{\tiny\it
eff}})$ is independent of $p$, and hence there is always a solution for which
the $(2n+1)$ quantities $\tilde{h}_{p}$ are equal, and positive. Henceforth,
let $\tilde{h}_{p}=\tilde{h}\;\forall p$. This physical solution is given by
the quadratic equation
\begin{equation}
\label{quadratic}
  \tilde{h}^2\left(2\sum_{q}
  (\beta_{pqqp}^{\mbox{\tiny\it eff}}+\beta_{pqpq}^{\mbox{\tiny\it eff}})
  \right) + \alpha\tilde{h}-1=0 \;\;\;\;\;\;\forall\: p.
\end{equation}
The coefficient of $\tilde{h}^2$ is now calculated.

  Firstly,
\begin{displaymath}
  \sum_{q}\beta_{pqqp}^{\mbox{\tiny\it eff}} = \frac{2\pi u}{\cal A}\!\int\!\!
  \phi_{p}^*\phi_{p}
  \sum_{q}\phi_{q}^*\phi_{q}d\Omega - \frac{8\pi^2n^2C^2}{{\cal A}^2\kappa}
  \sum_{l=2}^{2n}\sum_{m}\frac{(l+2)(l-1)}{l(l+1)}
  \int\!\!\phi_{p}^*\phi_{p}Y_{l}^m
  d\Omega \int\!\! Y_{l}^{m*}\sum_{q}\phi_{q}^*\phi_{q}d\Omega.
\end{displaymath}
The function $\sum_{q=-n}^n|\phi_{q}|^2$, which appears twice in the above
equation, may be written as
\begin{displaymath}
  \frac{(2n+1)!}{4\pi}\sum_{r=0}^{2n}\frac{\sin^{2r}\left(\frac{\theta}{2}
  \right)\cos^{2(2n-r)}\left(\frac{\theta}{2}\right)}{r!(2n-r)!}
\end{displaymath}
which is in the form of a binomial expansion of
\mbox{$(\sin^2(\frac{\theta}{2})+\cos^2(\frac{\theta}{2}))$}. It is therefore
just a constant; $\sum_{q}|\phi_{q}|^2=(2n+1)/4\pi$; and hence orthogonal to
all non-s-wave spherical harmonics. It follows that
\begin{displaymath}
  \sum_{q}\beta_{pqqp}^{\mbox{\tiny\it eff}}
  = (n+\mbox{$\frac{1}{2}$})\frac{u}{\cal A}.
\end{displaymath}

  Secondly, the sum $\sum_{q}\beta_{pqpq}^{\mbox{\tiny\it eff}}$ is found. This
also has two parts, deriving from equation (\ref{betaeff}), and the first has
already been calculated, since $\beta_{pqpq}=\beta_{pqqp}$. However, the order
of indices is relevant to the second part, which does not vanish this time.
There is no obvious method for simplifying this part, so the integrals have
been calculated as required for each value of $n$ of interest, with the result
that
\begin{displaymath}
  \sum_{q}(\beta_{pqqp}^{\mbox{\tiny\it eff}}
  +\beta_{pqpq}^{\mbox{\tiny\it eff}}) = (2n+1)\frac{u}{\cal A}
  - \frac{8\pi n^2C^2}{{\cal A}^2\kappa} f(n)
\end{displaymath}
where $f(1)=\frac{1}{12}$, $f(2)=\frac{43}{120}$, $f(4)=\frac{5471}{5040}$ and
$f(6)=\frac{1376527}{720720}$. So the solution to equation (\ref{quadratic}) is
\begin{equation}
\label{htilde}
  \frac{\varepsilon\tilde{h}}{2\omega}\approx -\frac{\alpha\omega}{\varepsilon}
  +\sqrt{\frac{\alpha^2\omega^2}{\varepsilon^2}+1}\;\;\;\;\;\;\forall\;p.
\end{equation}
where
\begin{displaymath}
  \varepsilon \equiv 4\sqrt{n+\mbox{$\frac{1}{2}$}-16\pi f(n)\mu}.
\end{displaymath}
As before, $\mu$ is the `malleability', $n^2C^2/{\cal A}u\kappa$, and $\omega$
is the `scale parameter', $\sqrt{{\cal A}/u}$.
From the definitions, the magnitude of the order parameter is
\begin{displaymath}
  \int|\psi|^2d{\cal A} = \sum_{p} \tilde{h}_{p} = (2n+1) \tilde{h}.
\end{displaymath}
In figure \ref{order}, a graph of $\varepsilon\tilde{h}/\omega$ against
$\alpha\omega/\varepsilon$ is plotted in bold. Recall that $\alpha$ is a
temperature-like variable, measured with respect to the shifted mean-field
transition temperature $r_{c}$. The figure shows that, within the lowest Landau
level, and Hartree-Fock approximation schemes, the transition is removed.
Instead there is a gradual change from a low-temperature region with a high
degree of order, to a poorly-ordered high temperature region, and this picture
may be closer to the truth. Of course, the Hartree-Fock approximation should
only be trusted at high temperature ($\alpha\omega/\varepsilon\gg 1$), as it is
rigorously correct to ${\cal O}(u)$. At low temperature, equation
(\ref{htilde}) tends asymptotically to
\mbox{$\tilde{h}=-\alpha\omega^2/(n+\mbox{$\frac{1}{2}$}-16\pi f(n)\mu)$},
which is at variance with the correct mean-field expression
\mbox{$\tilde{h}\rightarrow -\alpha/6\beta^{\mbox{\tiny\it eff}}_{0000}$}. For
$n=1$, the former expression becomes \mbox{$-3\alpha\omega^2/(18-16\pi\mu)$}
and the latter \mbox{$-5\alpha\omega^2/(18-8\pi\mu)$}. Nonetheless, the
Hartree-Fock expression is a useful qualitative indicator of the system's
behaviour. For comparison, the correct, lowest-Landau-level solution for $n=1$,
as calculated in section \ref{n=1}, is plotted as dashed lines in figure
\ref{order}, for various values of the malleability $\mu$. Note that some
$\mu$-dependance is contained in the function $\varepsilon$, which scales the
axes of the figure. It is apparent from the graphs that the Hartree-Fock
solution is correct at high temperature, and fairly poor at low temperature,
except for the special case when $\mu=9/14\pi$, for which the asymptotic
gradient is in agreement with the mean-field value. 

  The ordinate of figure \ref{order} is a measure of the order parameter, under
some normalization, while the abscissae is the temperature parameter $\alpha$
in units of $\varepsilon/\omega$. Hence it is immediately apparent for the
expression
for $\omega$ that the crossover between high- and low-temperature regimes
is sharper on larger vesicles. The crossover apparently becomes infinitely
sharp at a finite value of the malleability, when
\mbox{$\mu=\mu_{c}=(2n+1)/(32\pi f(n))$}. Once again, it is apparent, from
reference to the correct values of $\mu_{c}$ calculated in section
\ref{stability}, that the Hartree-Fock approximation gives
qualitatively correct, but quantitatively inaccurate results.

  The shape expectation values can now be found. Substituting for
$\left<a_{p}^*a_{q}\right>$ in equation (\ref{rho1}) with
$\tilde{h}\delta_{pq}$ gives
\begin{displaymath}
  \left<\rho_{lm}\right>=\frac{-\tilde{h}}{2\Delta_{l}}\sum_{p}\gamma_{pplm}.
\end{displaymath}
But $\sum_{p}\gamma_{pplm}=0$. So all spherical harmonic amplitudes have a
mean value of zero, despite the distorting influence of the $2n$ topological
defects in $\psi$. This is because all vortex positions have been thermally
averaged. The second moments are found, via equation (\ref{rho2}), from the
fourth moments of the order parameter field. From Wick's theorem,
\mbox{$\left<a_{p}^*a_{q}^*a_{r}a_{s}\right>=(\delta_{pr}\delta_{qs}+
\delta_{ps}\delta_{qr})\tilde{h}^2$}. So equation (\ref{rho2}) becomes
\begin{displaymath}
  \left<\rho_{l_{1}m_{1}}^*\rho_{l_{2}m_{2}}\right>
  =\frac{\tilde{h}^2}{4\Delta_{l_{1}}
  \Delta_{l_{2}}}\sum_{pq}\gamma_{qpl_{1}m_{1}}\gamma_{qpl_{2}m_{2}}
  +\frac{\delta_{l_{1}l_{2}}\delta_{m_{1}m_{2}}}{\Delta_{l_{1}}}.
\end{displaymath}
For instance, the auto-correlation function $\left<|\rho_{2 0}|^2\right>$ for
$n=1$ is evaluated as
\begin{displaymath}
  \left<|\rho_{20}|^2\right>=\frac{1}{12\kappa}\left(\frac{4\pi\mu}{5\omega^2}
  \tilde{h}^2+1\right).
\end{displaymath}
Hence deformations increase with the magnitude of the order parameter, as
expected.

\section{Conclusion}

  A new, simple expression has been derived for the coupling between the lowest
Landau levels of intrinsic $n$-atic order and the shape of a genus-zero
membrane. It has been shown, within the lowest-Landau-level approximation
scheme, that all spherical harmonic amplitudes of deformations of such a
vesicle have zero thermal average, but that their mean squares have a
contribution from deformation by the order, and a contribution from thermal
excitation.  Only a finite number of spherical harmonics are coupled to the
lowest Landau levels of order. It has been established that this order is
capable of fundamentally altering, and perhaps bursting a vesicle of finite
rigidity. This occurs at a critical value of the `malleability', for which the
model is strongly analogous to that of a marginal type-I-type-II
superconductor. As this critical value is approached, mean-field theory becomes
increasingly accurate. The model for fluid vesicles with vector order has been
solved exactly, using no further approximations, and the general result is
established that the crossover from high- to low- temperature behaviour becomes
sharper for larger vesicles. The solution also confirms that mean-field theory
becomes increasingly accurate as the critical malleability is approached. The
counter-intuitive result is ascertained that the two topological defects in the
order are more likely to be found far apart than close together, even at high
temperature, as the phase space volume element increases with defect
separation. The Hartree-Fock approximation is valid at high temperature and
significantly inaccurate at low temperature, but exhibits qualitatively
correct behaviour.

\section{Acknowledgements}

  Many thanks to Mike Moore for his contribution to this research. The work was
funded by the EPSRC; award reference number 92310214.

\appendix

\section{Solution of the Partition Function Double Integral}
\label{integral1}

  The integral required in equation (\ref{Z}) is
\begin{displaymath}
  I=\int_{-1}^1\!\!d\chi(1-\chi)\int_{0}^{\infty}\!\!t^5 e^{-(At^2+Bt^4)}dt
\end{displaymath}
where $A=a(\chi+3)$ and $B=b(\chi^2+10\chi+13)-c(\chi^2+3)$. Expanding the
first exponential as a sum and using
\mbox{$\int_{0}^{\infty}t^{2m+5}e^{-Bt^4}dt
=\Gamma(\frac{m+3}{2})/4B^{\frac{m+3}{2}}$} gives
\begin{displaymath}
  I=\mbox{$\frac{1}{4}$}\sum_{m=0}^{\infty} \frac{(-a)^m}{m!}
  \Gamma(\frac{m+3}{2}) \int_{-1}^{1}\frac{(1-\chi)(\chi+3)^m}
  {B^{\frac{m+3}{2}}} d\chi.
\end{displaymath}
Using the indefinite integral
\begin{displaymath}
  \int_{-1}^1 \frac{(1-\chi)(\chi+3)^{2n-1}}{B^{n+1}}d\chi
  = \frac{-(3+\chi)^{2n}}{2n(2b+3c)B^n} + \mbox{const.}
\end{displaymath}
gives
\begin{displaymath}
  I=\mbox{$\frac{1}{4}$} \sum_{m=0}^{\infty} \frac{(-a)^m(\frac{m+1}{2})!}
  {(m+1)!(2b+3c)} \left[\frac{1}{(b-c)^{\frac{m+1}{2}}} - \left(\frac{4}{6b-c}
  \right)^{\frac{m+1}{2}} \right].
\end{displaymath}
The exact form of the function $B(\chi)$ is crucial to the solubility of
the indefinite integral above. In particular, it is not easy to invent new
fields coupled to the dynamical variables to aid computation of their
expectation values. Now, applying the identity
\mbox{$(\frac{m}{2})!/m!\equiv \sqrt{\pi}/2^m(\frac{m-1}{2})!$} allows the
infinite sum to be separated into the standard forms of exponentials and
confluent hypergeometric functions thus:
\begin{eqnarray*}
  I=\frac{1}{16(2b+3c)} \left[
  \frac{2\sqrt{\pi}}{\sqrt{b-c}}\exp\left(\frac{a^2}{4(b-c)}\right) \right. &-&
  \frac{4\sqrt{\pi}}{\sqrt{6b-c}}\exp\left(\frac{a^2}{6b-c}\right)  \\
  + \left( \frac{8a}{6b-c}\right)\;F_{1}\!\!\!\!\!\!\!\!_{1}\;\;\:
  \left(1,\frac{3}{2};\frac{a^2}{6b-c}\right) &-& \left.
  \left(\frac{2a}{b-c}\right)\;F_{1}\!\!\!\!\!\!\!\!_{1}\;\;\:
  \left(1,\frac{3}{2};\frac{a^2}{4(b-c)}\right) \right].
\end{eqnarray*}
The identities \mbox{$\;F_{1}\!\!\!\!\!\!\!\!_{1}\;\;\:(a,c;x)
\equiv e^x\;F_{1}\!\!\!\!\!\!\!\!_{1}\;\;\:(c-a,c;-x)$} and
\mbox{$\mbox{\,erf\,}(x)\equiv
\frac{2x}{\sqrt{\pi}}\;F_{1}\!\!\!\!\!\!\!\!_{1}\;\;\:
(\mbox{$\frac{1}{2}$},\frac{3}{2};-x^2)$}
are used \cite{Arfken85} to give the result
\begin{displaymath}
  I=\frac{\sqrt{\pi}}{4a(2b+3c)} \left[ x\,e^{x^2}\!\mbox{\,erfc\,}{x}
  \right]^{\frac{a}{\sqrt{4b-4c}}}_{x=\frac{a}{\sqrt{6b-c}}}
\end{displaymath}
where $\mbox{\,erfc\,}$ is the complementary error function:
$\mbox{\,erfc\,}{x}=1-\mbox{\,erf\,}{x}$.

\section{Solution of the Constrained, Two-Point, Spherical Surface Integral}
\label{integral2}

  The integrals required for thermally averaging combinations of $a_{p}$ over
states of a given $t$ and $\chi$ are of the form
\begin{displaymath}
  J = \int g(\cos\theta_{1},\cos\theta_{2}) \,
  \delta(\mbox{\boldmath $n$}_{1}\cdot\mbox{\boldmath $n$}_{2}-\chi)
  d\Omega_{1} d\Omega_{2}
\end{displaymath}
where $g(x,y)$ is some function. The integral is over the
positions of two points on a unit sphere. The unit vectors
$\mbox{\boldmath $n$}_{1}$ and
$\mbox{\boldmath $n$}_{2}$ point from the centre of the sphere to the points,
each of which is
varied over the surface, in such a way that
$\mbox{\boldmath $n$}_{1}\cdot\mbox{\boldmath $n$}_{2}$ has the
constant value $\chi$. In spherical polar coordinates, the solid angle elements
are $d\Omega_{k}=d(\cos\theta_{k})d\varphi_{k}$. Hence
\begin{eqnarray*}
  J=\int_{-1}^1\!\!d(\cos\theta_{1}) & & \int_{-1}^1 \!\! d(\cos\theta_{2})\,
  g(\cos\theta_{1},\cos\theta_{2})\, \\
  & & \int_{0}^{2\pi} \!\! d\varphi_{1}\,
  \int_{0}^{2\pi}\!\! d(\varphi_{1}-\varphi_{2})\, \delta(\cos\theta_{1}
  \cos\theta_{2}+\sin\theta_{1}\sin\theta_{2}\cos(\varphi_{1}-\varphi_{2})
  -\chi)
\end{eqnarray*}
and
\begin{eqnarray*}
  \int_{0}^{2\pi}\!\! d\varphi\, \delta(\cos\theta_{1}\cos\theta_{2}
  &+& \sin\theta_{1}\sin\theta_{2}\cos\varphi-\chi) \\
  &=& 2\int_{-1}^1\!
  \frac{d(\cos\varphi)}{\sin\theta_{1}\sin\theta_{2}\sin\varphi} \delta(
  \cos\theta_{1}\cos\theta_{2}+\sin\theta_{1}\sin\theta_{2}\cos\varphi-\chi) \\
  &=& \frac{2 \Theta\left(\left(\frac{\chi-\cos\theta_{1}\cos\theta_{2}}
  {\sin\theta_{1}\sin\theta_{2}}\right)^2\right)}
  {\sqrt{\sin^2\theta_{1}\sin^2\theta_{2}-(\chi-\cos\theta_{1}\cos\theta_{2}
  )^2}}
\end{eqnarray*}
where $\Theta(x)$ is the Heaviside step function; $\Theta(x<0)=0$,
$\Theta(x>0)=1$. Now, setting $x=\cos\theta_{1}$ and $y=\cos\theta_{2}$ gives
\begin{displaymath}
  J=4\pi\int_{-1}^1 \!\! dx \int_{-1}^1 \!\! dy \frac{g(x,y)}
  {\sqrt{(1-x^2)(1-y^2)-(\chi-xy)^2}}\,\Theta\left( \frac{(\chi-xy)^2}
  {(1-x^2)(1-y^2)}\right).
\end{displaymath}
The region of the $x$-$y$ plane for which the Heaviside function has a positive
argument is the interior of an ellipse, contained within the unit square of
integration. The transformation
\begin{eqnarray*}
  2x &=& (\sqrt{1+\chi}+\sqrt{1-\chi})u+(\sqrt{1+\chi}-\sqrt{1-\chi})v \\
  2y &=& (\sqrt{1+\chi}-\sqrt{1-\chi})u+(\sqrt{1+\chi}+\sqrt{1-\chi})v
\end{eqnarray*}
with Jacobian
\begin{displaymath}
  \frac{\partial(x,y)}{\partial(u,v)} = \sqrt{1-\chi^2}
\end{displaymath}
maps the ellipse onto a unit circle in the $u$-$v$ plane thus:
\begin{displaymath}
  J=4\pi \int_{-1}^1 \!\! du\, \int_{-\sqrt{1-u^2}}^{\sqrt{1-u^2}}\!\! dv
  \frac{g(x,y)}{\sqrt{1-u^2-v^2}}.
\end{displaymath}
Hence a final transformation to polar coordinates $(s,\xi)$ gives
\begin{displaymath}
  J = 4\pi\int_{0}^{1}\!\!\!ds \int_{0}^{2\pi}\!\!\!d\xi
  \frac{s}{\sqrt{1-s^2}}\,g(x,y) 
\end{displaymath}
with
\begin{eqnarray*}
  x &=& \mbox{$\frac{1}{2}$} s \left[ (\sqrt{1+\chi}+\sqrt{1-\chi})\cos\xi
  + (\sqrt{1+\chi}-\sqrt{1-\chi})\sin\xi \right] \\
  y &=& \mbox{$\frac{1}{2}$} s \left[ (\sqrt{1+\chi}-\sqrt{1-\chi})\cos\xi
  + (\sqrt{1+\chi}+\sqrt{1-\chi})\sin\xi \right].
\end{eqnarray*}
So, for instance,
\begin{displaymath}
  \int\!\! \delta(\mbox{\boldmath $n$}_{1}\cdot\mbox{\boldmath $n$}_{2}-\chi)\,
  d\Omega_{1}d\Omega_{2} = 8\pi^2
\end{displaymath}
and
\begin{displaymath}
  \int\!\! \cos\theta_{1}\cos\theta_{2}
  \delta(\mbox{\boldmath $n$}_{1}\cdot\mbox{\boldmath $n$}_{2}-\chi)\,
  d\Omega_{1}d\Omega_{2} = \frac{8\pi^2\chi}{3}.
\end{displaymath}

\begin{figure}
  \caption{Graph of the expectation value of the cosine of the angle subtended
at the vesicle's centre by the two vortices $\left<\chi\right>$ against the
temperature-like combination $\alpha\omega$, for the cases $\mu=0$ (the rigid
sphere), $\mu=\mu_{c}$ (critical malleability), and $\mu=\mu_{c}/2$. The
vortices are antipodal at low temperature, and tend to be well
separated even at high temperature due to the shape of the phase space.}
  \label{chi}
\end{figure}

\begin{figure}
  \caption{Graph of the normalized order parameter
$\frac{(18-8\pi\mu)}{15\omega}\left<\int|\psi^2|d{\cal A}\right>$ against the
temperature-like quantity $\alpha\omega$, for various values of the
malleability $\mu$. As $\mu$ approaches the critical value $\mu_{c}=9/4\pi$,
low-temperature order increases, since the vesicle is intrinsically flattened,
but high-temperature order is independent of $\mu$. Hence mean-field theory
becomes accurate in the marginal type-II-type-I limit, $\mu=\mu_{c}$.}
  \label{order1}
\end{figure}

\begin{figure}
  \caption{Graph of the mean-square amplitude of the $Y_{2}^0$ mode of
deformation multiplied by $(\mu-\mu_{c})^2\kappa$, against the temperature-like
quantity $\alpha\omega$, for the cases $\mu=\mu_{c}/2$ and $\mu=\mu_{c}$.}
  \label{deform}
\end{figure}

\begin{figure}
  \caption{Graph of $\varepsilon\tilde{h}/\omega$ against
$\alpha\omega/\varepsilon$, showing the non-singular progression from the
low-temperature, highly ordered state to the high temperature poorly ordered
state. The Hartree-Fock solution is show as a continuous, bold line. For
comparison, the more accurate solution for $n=1$, as calculated
non-perturbatively in section \ref{n=1}, is shown dashed, for various values of
the malleability $\mu$.}
  \label{order}
\end{figure}


\begin{thebibliography}{999}

\bibitem{Lipowsky91} R. Lipowsky, Nature {\bf 349}, 475 (1991).
\bibitem{David89} F. David, Geometry and Field Theory of Random Surfaces and
Membranes, in {\em Statistical Mechanics of Membranes and Surfaces}, ed.
D. Nelson, T. Piran and S. Weinberg (World Scientific Publishing Co.\ Pte.\
Ltd., 1989).
\bibitem{Peliti94} L. Peliti, {\em Fluctuations of Membranes},
cond-mat/9501076.
\bibitem{MacKintosh91} F. C. MacKintosh and T. C. Lubensky, Phys.\ Rev.\
Lett. {\bf 67} 1169 (1991).
\bibitem{Nelson79} D. R. Nelson and B. I. Halperin, Phys.\ Rev.\ B {\bf 19},
2457 (1979).
\bibitem{Evans95} R. M. L. Evans, J. Phys.\ II (France) {\bf 5} 507 (1995).
\bibitem{Helfrich86} W. Helfrich, J. Phys.\ (Paris) {\bf 47}, 321 (1986).
\bibitem{flat} E. A. Abbott, {\em Flatland: A Romance of Many Dimensions}
(Dover, New York, 1952).
\bibitem{Rutherford59} D. E. Rutherford, {\em Vector Methods} (Oliver \& Boyd
Ltd., 1959).
\bibitem{Weeks85} Jeffrey R. Weeks, {\em The Shape of Space} (Marcel Dekker,
1985).
\bibitem{Park92} J. Park, T. C. Lubensky and F. C. MacKintosh, Europhys.\
Lett.\ {\bf 20,3} 279 (1992).
\bibitem{Landau58} L. D. Landau and E. M. Lifshitz, {\em Quantum Mechanics:
Non-Relativistic Theory} (Oxford Pergamon, 1958).
\bibitem{Abrikosov} A. A. Abrikosov, Sh.\ Eksp.\ Teor.\ Fiz.\ {\bf 32}, 1442
(1957) [Sov.\ Phys.\ -JETP {\bf 5}, 1174 (1957)].
\bibitem{Roy83} S. M. Roy and Virenda Singh, Phys.\ Rev.\ Lett.\ {\bf 51} 2069
(1983).
\bibitem{ONeill93} J. A. O'Neill and M. A. Moore, Phys.\ Rev.\ B {\bf 48} 374
(1993).
\bibitem{Cai94} W. Cai, T. Lubensky, P. Nelson, and T. Powers, ``Measure
factors, tension, and correlations of fluid membranes,'' J. Phys.\ II France
{\bf 4} (1994) 931--949.
\bibitem{Ruggeri76} See for instance G. J. Ruggeri and D. J. Thouless, J. 
Phys.\ F {\bf 6} 2063 (1976).
\bibitem{Ma} See for instance S. K. Ma, {\em Modern Theory of Critical
Phenomena} (W. A. Benjamin, Inc., 1976).
\bibitem{Arfken85} G. Arfken, {\em Mathematical Methods for Physicists, third
edition} (Academic Press Ltd., 1985).

\end{thebibliography}
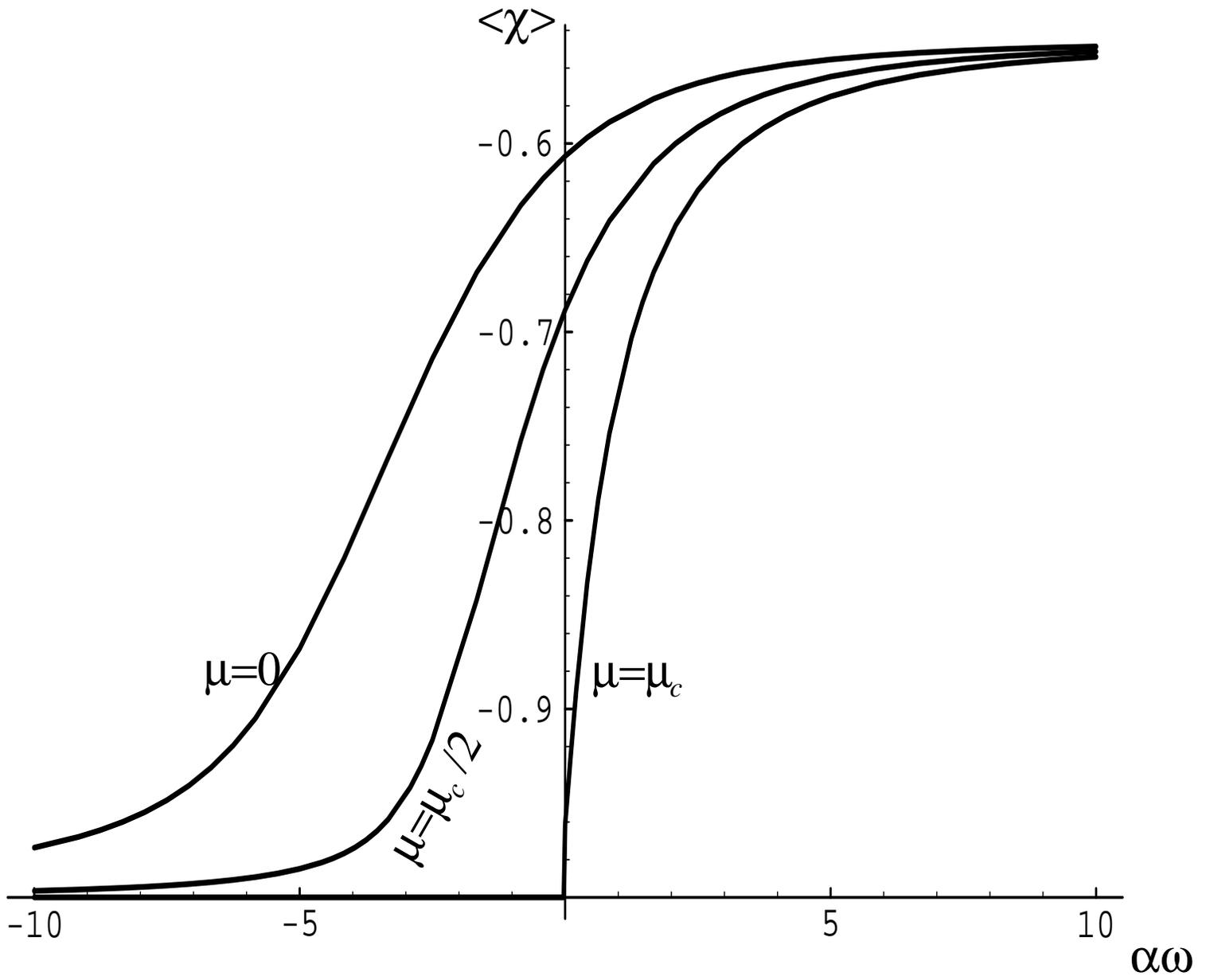
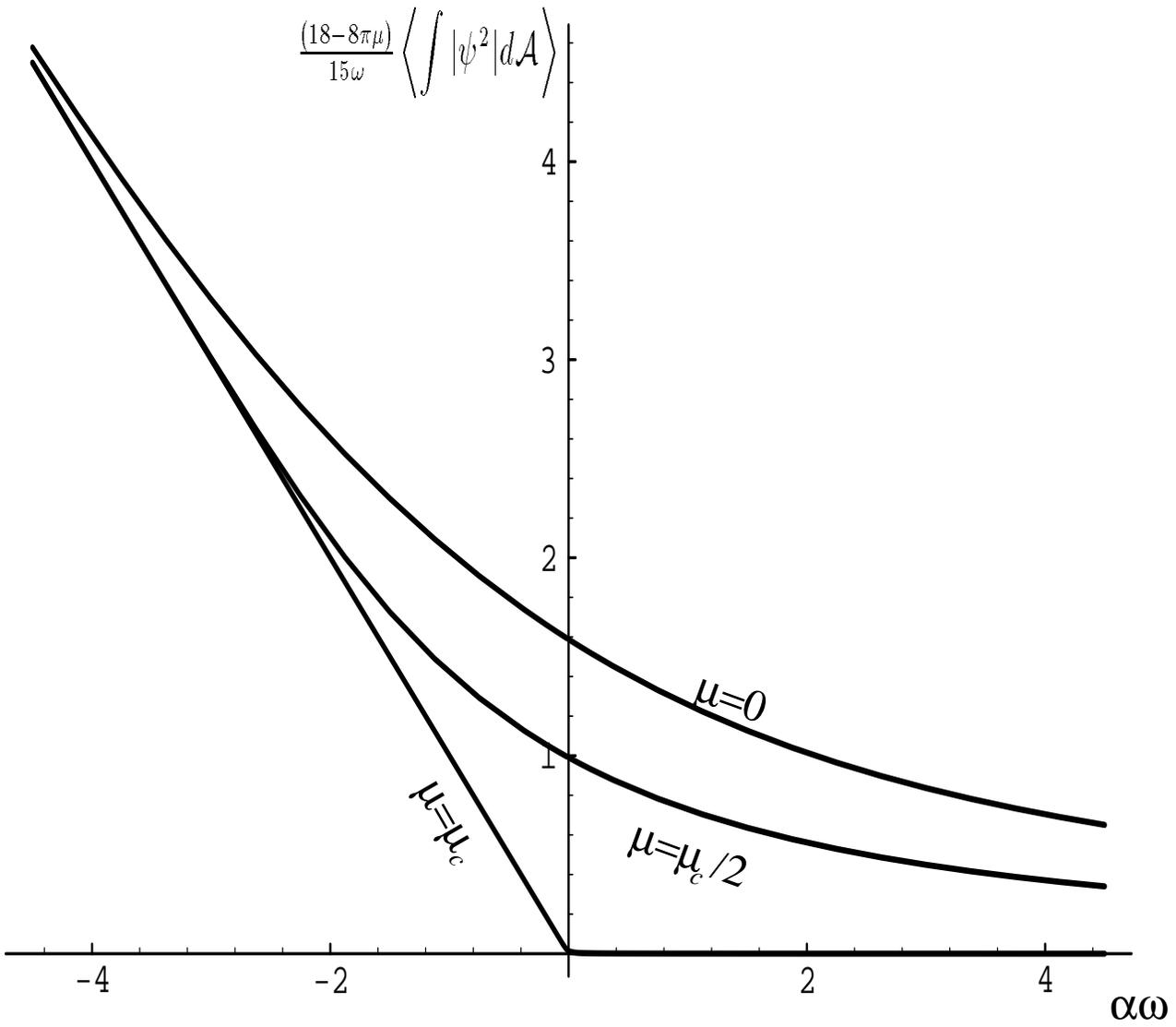
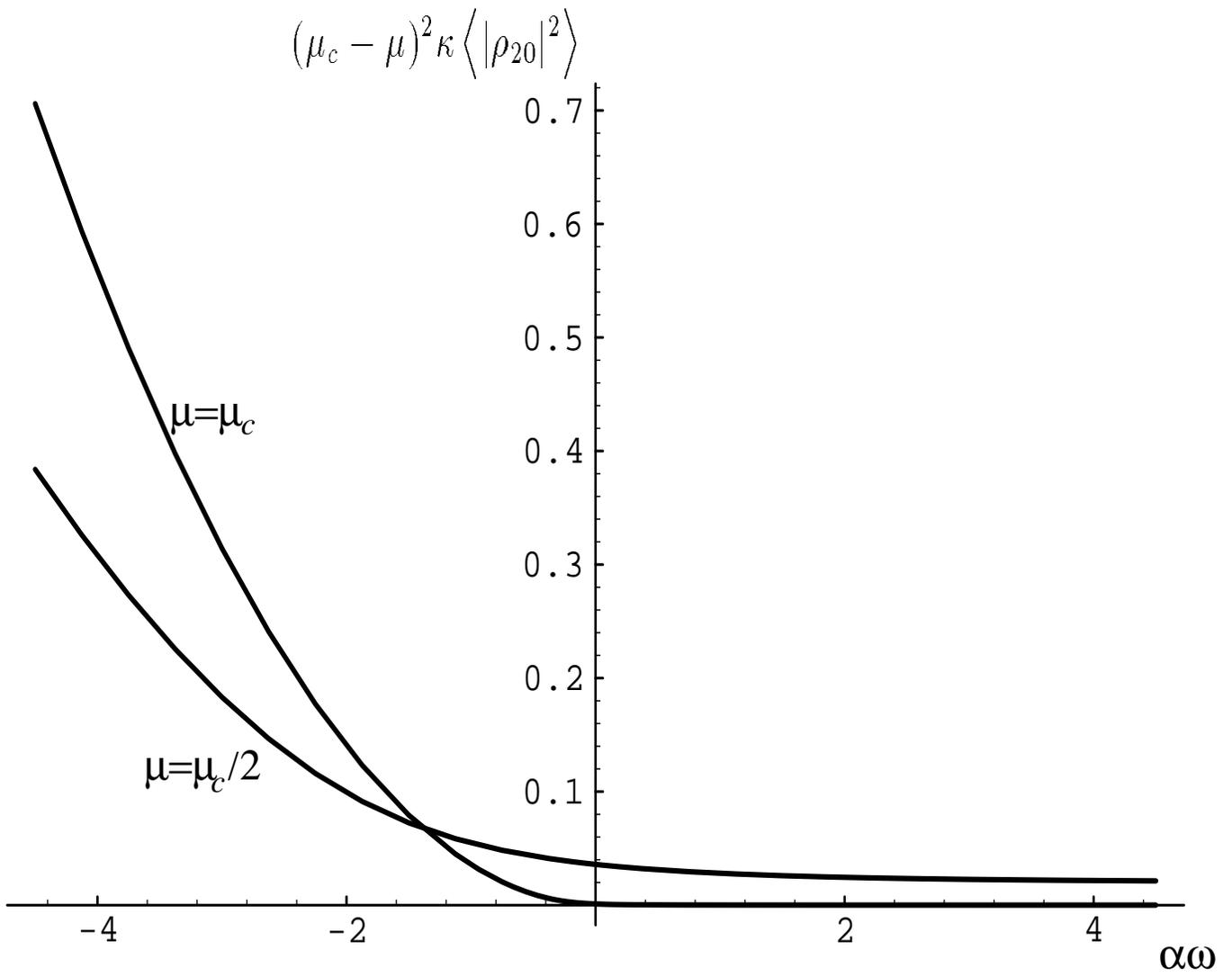
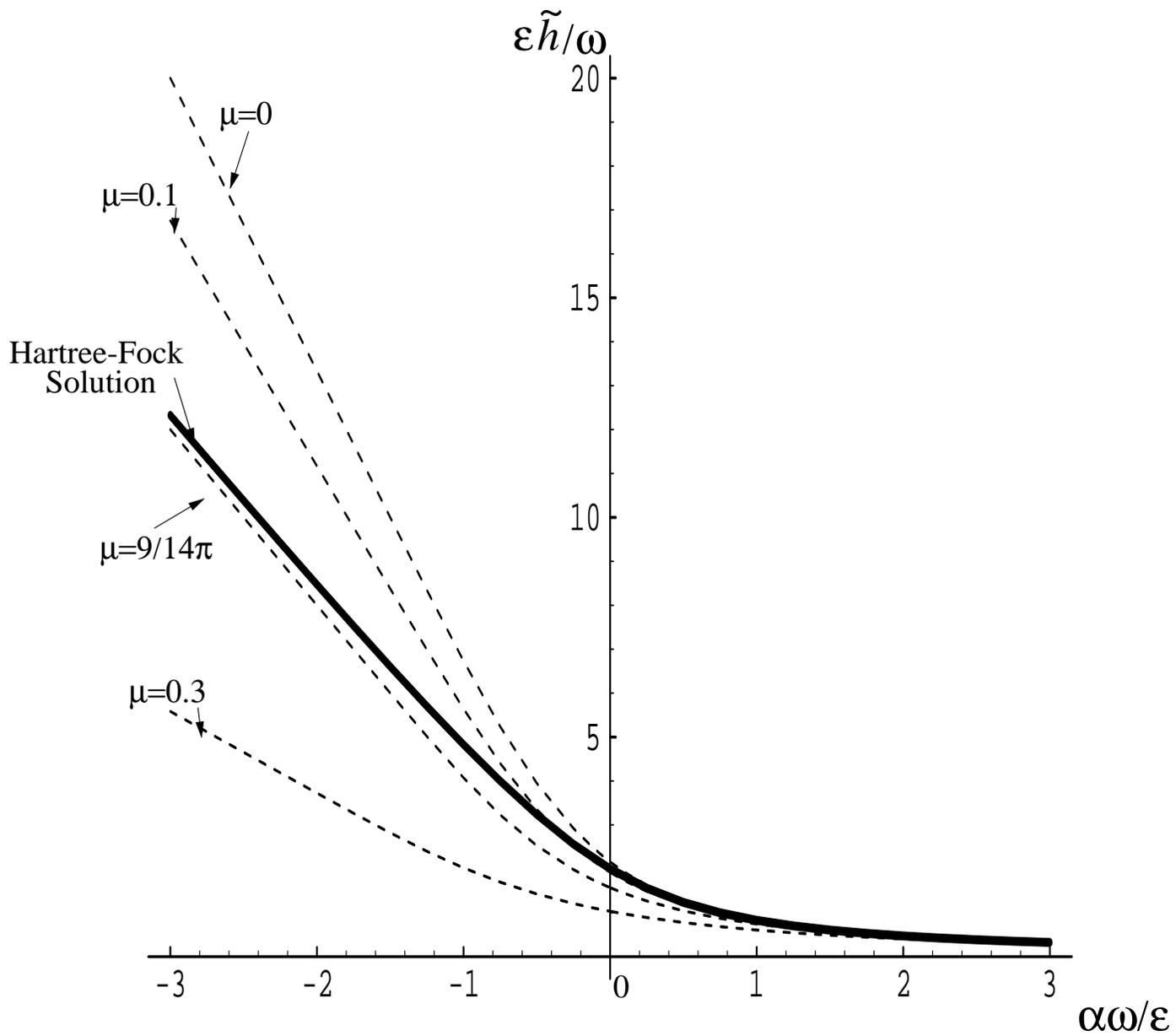
\end{document}